%% file: paper.tex
\documentclass[letterpaper,twocolumn,10pt]{article}
\usepackage{usenix-2020-09}

\usepackage{tikz}
\usepackage{amsmath,amssymb}
\usepackage[normalem]{ulem}

\usepackage{filecontents}
\usepackage{listings}
\usepackage{xspace}
\usepackage{graphicx}
\usepackage{caption}
\usepackage[export]{adjustbox}
\usepackage{subcaption}
\usepackage{balance}
\usepackage{epsfig}
\usepackage{multirow}
\usepackage{url}
\usepackage{pgf}
\usepackage{cleveref}
\usepackage{pifont}
\usepackage{booktabs}
\usepackage[compact]{titlesec}
\usepackage{hyperref}
\usepackage{theorem}
\usepackage{enumitem}
\usepackage{threeparttable}

\input{commands}
\begin{document}

\date{}

\title{\Large \bf \spk: Economizing Self-Healing LLM Training at Scale}

\author{
{\rm Tao He\textsuperscript{1}, 
Xue Li\textsuperscript{1}, 
Zhibin Wang\textsuperscript{1,2}, 
Kun Qian\textsuperscript{1}, 
Jingbo Xu\textsuperscript{1}, 
Wenyuan Yu\textsuperscript{1}, 
Jingren Zhou\textsuperscript{1}}
\\
\\
\textsuperscript{1}Alibaba Group, 
\textsuperscript{2}Nanjing University
\\
\\
\href{mailto:unicron@alibaba-inc.com}{unicron@alibaba-inc.com}
} 
\maketitle

\begin{abstract}

Training large-scale language models is increasingly critical in various domains, but it is hindered by frequent failures, leading to significant time and economic costs. Current failure recovery methods in cloud-based settings inadequately address the diverse and complex scenarios that arise, focusing narrowly on erasing downtime for individual tasks without considering the overall cost impact on a cluster.

We introduce \spk, a workload manager designed for efficient self-healing in large-scale language model training. \spk~optimizes the training process by minimizing failure-related costs across multiple concurrent tasks within a cluster. Its key features include in-band error detection for real-time error identification without extra overhead, a dynamic cost-aware plan generation mechanism for optimal reconfiguration, and an efficient transition strategy to reduce downtime during state changes. Deployed on a 128-GPU distributed cluster, \spk~demonstrates up to a $1.9\times$ improvement in training efficiency over state-of-the-art methods, significantly reducing failure recovery costs and enhancing the reliability of large-scale language model training.

\end{abstract}

\input{sec-introduction}
\input{sec-background}

\input{sec-system}
\input{sec-detection}

\input{sec-model}

\input{sec-transition}
\input{sec-evaluation}

\input{sec-conclusion}

\Urlmuskip=0mu plus 1mu\relax
\balance
\bibliographystyle{plain}
\bibliography{paper}

\end{document}

%% file: commands.tex
\newcommand{\stitle}[1]{\vspace{0.5ex}\noindent{\bf #1}}

\newcommand{\kw}[1]{\textsf{#1}\xspace}
\newcommand{\spk}{{\kw{Unicron}}}
\newcommand{\benefit}{WAF}
\newcommand{\alibaba}{Alibaba Cloud}

\newcommand{\eat}[1]{}

\newcommand{\sev}[1]{\textsc{sev#1}}

\newsavebox{\bigimage}

\definecolor{codegreen}{rgb}{0,0.6,0}
\definecolor{codegray}{rgb}{0.5,0.5,0.5}
\definecolor{codered}{rgb}{0.59,0.0,0.09}
\definecolor{backcolour}{rgb}{0.96,0.96,0.96}
\definecolor{codeorange}{rgb}{1.0,0.2,0.0}
\definecolor{commentgreen}{RGB}{2,112,10}
\makeatletter
\def\UrlAlphabet{%
      \do\a\do\b\do\c\do\d\do\e\do\f\do\g\do\h\do\i\do\j%
      \do\k\do\l\do\m\do\n\do\o\do\p\do\q\do\r\do\s\do\t%
      \do\u\do\v\do\w\do\x\do\y\do\z\do\A\do\B\do\C\do\D%
      \do\E\do\F\do\G\do\H\do\I\do\J\do\K\do\L\do\M\do\N%
      \do\O\do\P\do\Q\do\R\do\S\do\T\do\U\do\V\do\W\do\X%
      \do\Y\do\Z}
\def\UrlDigits{\do\1\do\2\do\3\do\4\do\5\do\6\do\7\do\8\do\9\do\0}
\g@addto@macro{\UrlBreaks}{\UrlOrds}
\g@addto@macro{\UrlBreaks}{\UrlAlphabet}
\g@addto@macro{\UrlBreaks}{\UrlDigits}
\makeatother

\theoremstyle{definition}

\theoremstyle{remark}

\theoremstyle{property}

%% file: sec-introduction.tex
\section{Introduction}
\label{sec:introduction}
Large language models (LLMs) like ChatGPT~\cite{openai2021chatgpt}, BERT~\cite{devlin2018bert}, BLOOM~\cite{scao2022bloom}, Llama~\cite{touvron2023llama}, and Llama-2~\cite{touvron2023llama-2} are widely used in various real-world applications~\cite{devlin2018bert,radford2019language, jungherr2023using, chen2023llm}, 
drawing significant attention from both academia and industry for their role in advancing natural language processing and AI. 
These comprehensive models, often comprising billions of parameters, are trained on large-scale GPU clusters~\cite{shoeybi2019megatron,raffel2020exploring}.
To facilitate their training on thousands of GPUs, distributed frameworks like Megatron-LM (Megatron)~\cite{shoeybi2019megatron,narayanan2021efficient} and DeepSpeed~\cite{rasley2020deepspeed} have been developed, offering efficient parallelization and optimization.

\begin{figure}[t]
  \centering
  \vspace*{-4ex}
  \includegraphics[width=0.9\linewidth]{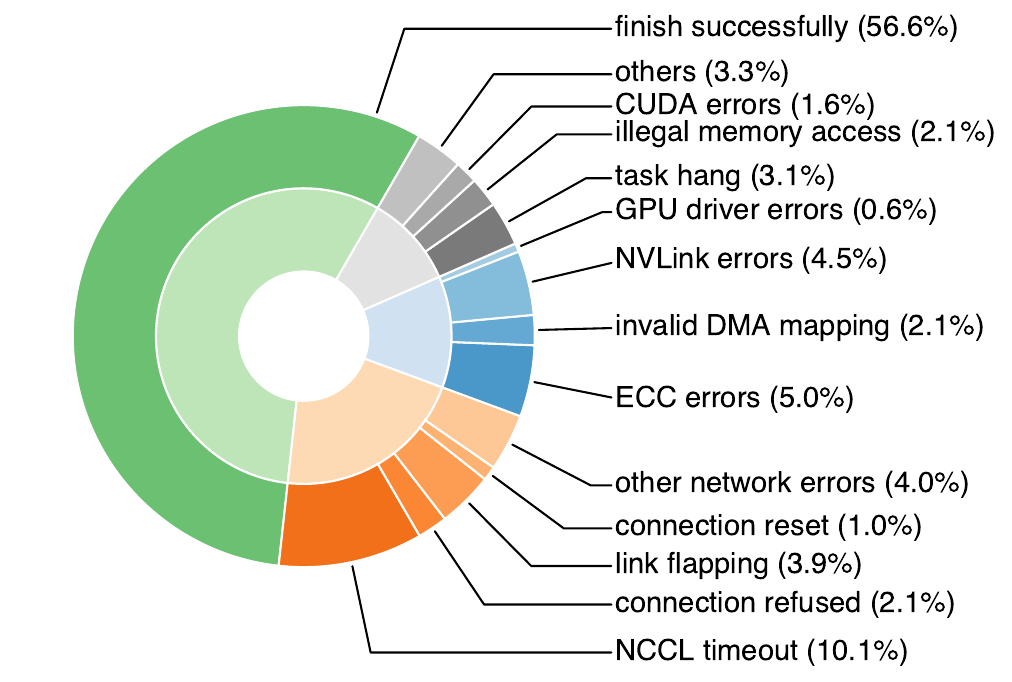}
  \caption{Distribution of task termination statistics.}
  \label{fig:errors}
  \vspace*{-2.5ex}
\end{figure}

With advanced models demanding extensive computational capabilities, cloud platforms provide a practical and cost-effective approach to Large Language Model (LLM) training by enabling the straightforward provisioning of GPU-rich clusters as required. Notable cloud providers such as Amazon Web Services~\cite{aws}, Google Cloud Platform~\cite{googlecloud}, Microsoft Azure~\cite{azure}, and Alibaba Cloud~\cite{alibabacloud} provide specialized services tailored to distributed model training.
However, training failures remain a common challenge, primarily due to the considerable volume of deployed resources and extended training durations, as underscored in several recent studies~\cite{wang2023gemini, jang2023oobleck, zhang2022opt, scao2022bloom, wu2023transom}. Our analysis of one of the major cloud platforms, referred to as \alibaba~for clarity in this paper, echoes this observation. We found that the failure rates for LLM training tasks can skyrocket to $43.4$\% for the top 5\% most resource-intensive tasks, as illustrated in Figure~\ref{fig:errors}.

\begin{figure}[t]
  \centering
  \includegraphics[width=\linewidth]{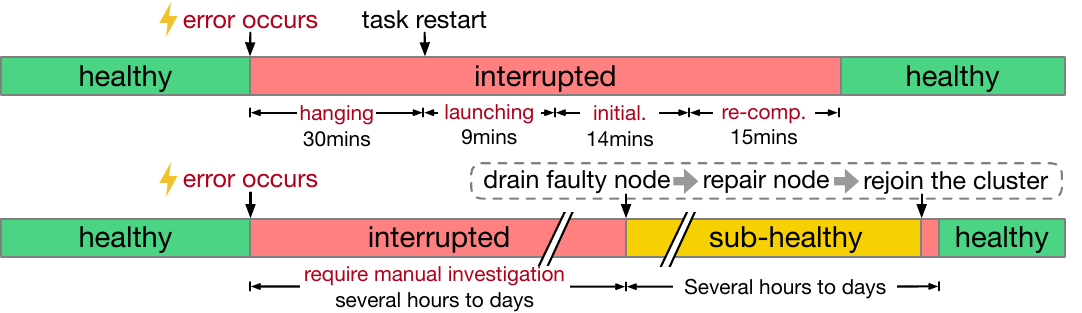}
  \caption{Training process with manual failure recovery.}
  \label{fig:stage}
  \vspace*{-2ex}
\end{figure}

Figure~\ref{fig:stage} delineates the divergent paths of recovery in the event of errors during a GPT-3 training exercise on \alibaba, leveraging a fleet of 256 NVIDIA H800 GPUs within the Megatron framework. For transient faults, which constitute 73\% of all errors and are typically remediable by restarting the system, the recovery trajectory may involve a system hang lasting up to 30 minutes -- stemming from the \texttt{all-reduce} communication timeout. This delay is followed by task termination and a succession of steps including a 9-minute wait for task resubmission, a 14-minute span for environment and CUDA configuration, and a final 15-minute recomputation phase. Collectively, this results in a downtime of 68 minutes. Conversely, hardware faults -- which precipitate the need for node drainage in 37\% of cases -- set off a more labor-intensive recovery entailing manual fault identification, node drainage, down-scaling of Megatron's configuration, and checkpoint realignment before resumption of training. This manual labor can extend the `interrupted' state for several hours to days, relegating the system to a `sub-healthy' state of compromised capacity. The scarcity and high cost of GPU resources exacerbate the financial stakes of these failures, as any inefficiency or idle time translates directly into considerable economic loss by squandering invaluable training time.

Despite these challenges, a range of methods have been developed to mitigate the impacts of training failures. However, existing methods often fail to provide comprehensive solutions, as they primarily focus on individual aspects of LLM training failure handling. For instance, some studies~\cite{eisenman2022check,wang2023gemini, mohan2021checkfreq, nicolae2020deepfreeze, wu2023transom} concentrate on the checkpointing process with the aim of reducing interruptions durations. Others propose elastic training and scheduling strategies~\cite{jang2023oobleck, KungFu, Elastic, 9373916, Elan, Lyra, shukla2022singularity, athlur2022varuna}, while additional works~\cite{thorpe2023bamboo,athlur2022varuna} explore the use of redundant computation or hot spares to prevent task interruptions. However, these solutions typically overlook the complexities of failures, which require an all-encompassing recovery strategy -- from quick error detection to swift recovery to a healthy state and seamless transitions between task running states when nodes are draining or joining. Consequently, there is a significant gap in the current methodologies which necessitates a more holistic approach.

\begin{figure}
  \centering
  \begin{subfigure}{0.9\linewidth}
    \centering
    \includegraphics[width=0.9\linewidth]{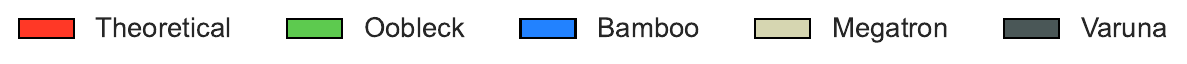}
  \end{subfigure}

  \centering
  \begin{subfigure}{0.48\linewidth}
    \centering
    \includegraphics[width=1.0\linewidth]{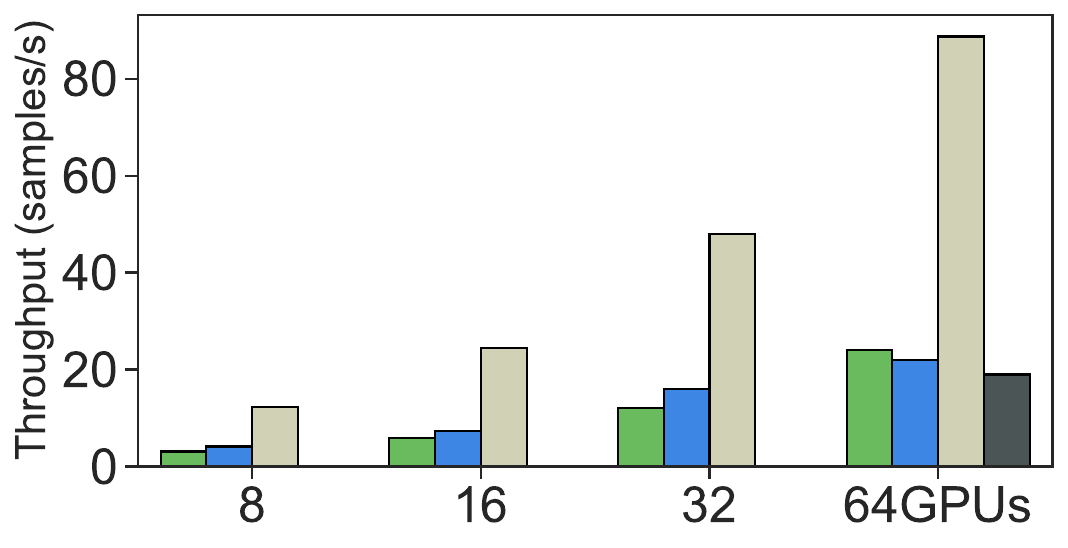}
    \caption{Throughput of training without failures.}
    \label{fig:training-throughput}
  \end{subfigure}
  \hspace{1ex}
  \begin{subfigure}{0.48\linewidth}
    \centering
    \includegraphics[width=1.0\linewidth]{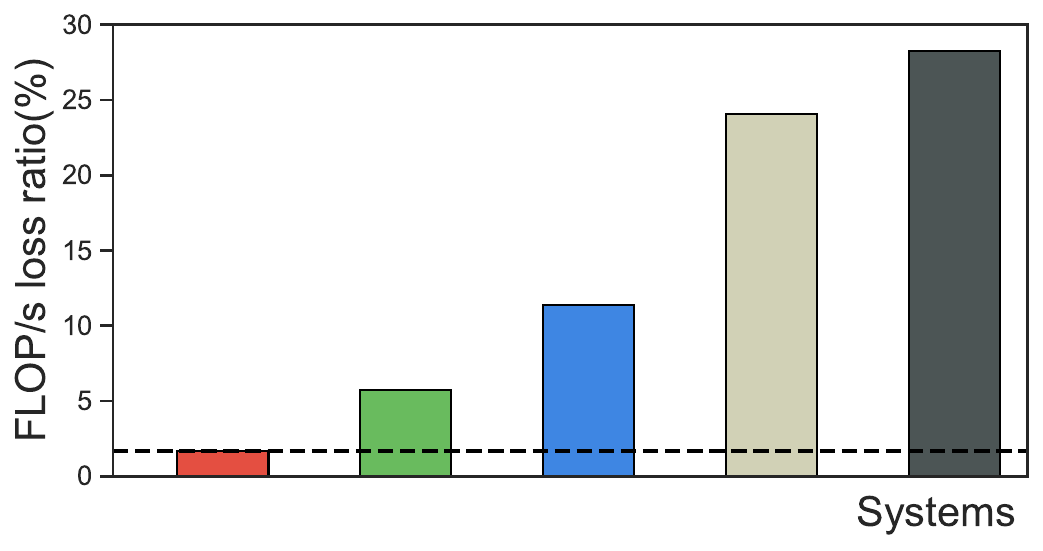}
    \caption{The FLOP/s reduction caused by failures.}
    \label{fig:flops-losses}
  \end{subfigure}

  \label{fig:throughput-and-errors}
  \caption{Throughput and FLOP/s reduction of training the GPT-3 7B model on a cluster of 64 GPUs for 7 days with 10 node fault errors occurred during the period. (a) The throughput is the number of samples the system can process per second. (b) The theoretical reduction is the ratio of hardware resources due to the unavailability during failures. For each system, the reduction is the percentage of lost FLOP/s compared with the ideal FLOP/s it can achieved assuming no failure happens.}
  \vspace*{-1ex}
\end{figure}

Moreover, systems specialized for training resilience such as Oobleck\cite{jang2023oobleck}, Bamboo\cite{thorpe2023bamboo}, and Varuna\cite{athlur2022varuna} operate at a fraction of Megatron's efficiency, leading to a scenario where resources are expended but not effectively utilized for training, as demonstrated in Figure~\ref{fig:training-throughput}. This inefficiency becomes even more pronounced when considering the throughput losses due to failures. As Figure~\ref{fig:flops-losses} reveals, a mere 2\% downtime can lead to throughput losses that are threefold or greater than the optimal scenario (compared with their own respective implementations). Such discrepancies indicate a misalignment in fault recovery objectives: {\em the essence lies not in sustaining training processes through failures but in economizing the entire training to minimize lost throughput}.
 
\stitle{\spk}. To address existing limitations in failure recovery during LLM training on cloud platforms, we introduce \spk, a distributed workload manager built with Megatron. \spk~is designed to enhance the training process by adopting a comprehensive strategy that focuses on minimizing the total cost of failures. This is achieved through a combination of efficient error detection, seamless transitions in system states, and effective management of sub-optimal conditions.

\spk's key features include inheriting all optimization techniques from Megatron, ensuring efficient training task execution. It preserves strict optimizer semantics, guaranteeing exact parameter updates without resorting to asynchronous or approximate methods during recovery from faults. The system's non-intrusive design facilitates easy adaptation to existing workloads and seamless integration with future updates to Megatron. Furthermore, \spk's self-healing capabilities enable it to efficiently detect and recover from a variety of failures. Its multi-tasking approach maximizes overall resource utilization across multiple tasks within a cluster, embodying economic efficiency in large-scale LLM training.
Our key contributions can be outlined as follows:
\begin{itemize}[noitemsep,topsep=2pt,leftmargin=*]\itemsep0em
  \item We strategized the architectural design of \spk~(Section~\ref{sec:system}), integrating it with Megatron~\cite{narayanan2021efficient}~for its high performance training capabilities and inserting new features to streamline failure recovery. 
  \item We developed efficient error detection and handling approaches (Section~\ref{sec:detection}), enabling \spk~to promptly identify failures during execution and initiate suitable corrective actions tailored to the particular failure causes.
  \item We formulated a cost-aware plan generation mechanism (Section~\ref{sec:model}), aiding \spk~in configuring the most optimal plan. This mechanism is informed by a model considering the multiplicity of tasks within the cluster.
  \item We introduced a transition strategy (Section~\ref{sec:transition}), which minimizes system transition durations by exploiting partial results from ongoing training iterations and leveraging the nearest principle for state migration. 
  \item We conducted extensive experiments, using a myriad of representative training tasks on a cluster comprised of 128 GPUs (Section~\ref{sec:evaluation}). These experiments revealed that \spk~markedly reduces costs associated with recovery from failures resulting in an enhancement of up to $1.9\times$ in overall training efficiency.
\end{itemize}

%% file: sec-background.tex
\section{Background and Opportunities}
\label{sec:background}

This section provides a comprehensive overview of the architecture underpinning distributed LLM training, delves into the prevailing statistics on system failures, and evaluates the current strategies implemented for failure recovery. Furthermore, it elucidates the opportunities that these challenges present, laying the groundwork for transformative approaches in resilience and efficiency within this domain.

\subsection{LLM Training}
\label{sec:background:training}

\stitle{Training Frameworks.} The scale of data and computational intensity required for training Large Language Models (LLMs) necessitates the adoption of distributed training frameworks. Recent benchmarks, such as the MLPerf results~\cite{mlperf}, showcase the formidable capabilities of NVIDIA's accelerated computing platforms. Specifically, leveraging 10,752 NVIDIA H100 Tensor Core GPUs coupled with Quantum-2 InfiniBand networking, NVIDIA has achieved the impressive feat of training a GPT-3 model with 175 billion parameters over 3.7 trillion tokens in just eight days~\cite{mlperf_v3.1}. The cornerstone of this achievement is Megatron~\cite{shoeybi2019megatron} -- a robust transformer architecture by NVIDIA that integrates a slew of advanced optimizations, ensuring over 50\% of FLOP/s utilization of available computational resources, a benchmark substantiated by Figure~\ref{fig:training-throughput}. This blend of performance and efficiency positions Megatron as a venerated choice for LLM training
\cite{scao2022bloom, glm, bai2023qwen}.

\stitle{Parallelism Approaches.} The quest for efficiency in distributed training has given rise to various parallelism strategies. Data Parallelism (DP)~\cite{li2020pytorch, hwang2021elastic, rasley2020deepspeed, karakus2021amazon} distributes the workload evenly across multiple workers, each processing a distinct subset of data. In contrast, Pipeline Parallelism (PP)~\cite{huang2019gpipe, narayanan2019pipedream, narayanan2021memory, kim2023bpipe, rasley2020deepspeed, fan2021dapple} slices the model into sequential stages, optimizing the process via micro-batches. Tensor Parallelism (TP)~\cite{narayanan2021efficient, shoeybi2019megatron}, another variant, opts for vertical model partitioning. The fusion of DP, PP, and TP—termed `3D Parallelism'—along with additional techniques like sequence parallelism~\cite{li2021sequence, korthikanti2023reducing} and gradient checkpointing~\cite{chen2016sublinear, jain2020checkmate}, enhances the adaptability and scalability of LLM training, affording a degree of natural elasticity in managing resources within the constraints of memory limitations.

\subsection{Failure Statistics}
\label{section:background:stats}
In the domain of LLM training, failures are an unwelcome yet common occurrence, often arising from the immense computational loads and extended training periods required. The root causes are varied, encompassing complexities in network topology and inconsistencies in hardware reliability~\cite{SageMaker}. An examination of the failure patterns on the \alibaba~platform has revealed that the most resource-intensive tasks—representing the top 5\% -- exhibit a startling 43.4\% rate of abnormal terminations, underscoring the critical need for more resilient training systems.

Contrary to what one might expect, these failures are not continuous but sporadic. The bulk of GPUs perform without issues for the majority of the time. In a typical set of 128 GPUs, failure frequencies range from once to seven times weekly, equating to an average time between failures of over a day. This pattern of infrequent yet impactful disruptions mirrors trends reported in Meta's training of the OPT model~\cite{zhang2022opt}.

\stitle{Failure Recovery Cost.}
The recovery from such failures, as depicted in Figure~\ref{fig:stage}, is a multi-stage process beginning with error detection and the determination of a viable configuration for resumption. The subsequent steps involve transitioning to this new configuration and then continuing training, potentially at a reduced efficiency. This process incurs distinct costs at each phase—whether managed manually or automatically—including the time lost during error detection, the effort required to reconfigure the system, and the diminished throughput in less-than-ideal operational states. These costs, which collectively impact the overall efficiency and output of the training process, can be formalized as follows:

\begin{equation}
  \label{eq:cost}
  C_{\text{recovery}} = C_{\text{detection}} + C_{\text{transition}} + C_{\text{sub-healthy}}
\end{equation}

\subsection{Related Work}

\stitle{Error Detection.}
Contemporary cloud services offer fundamental monitoring tools that allow for basic oversight of system operations~\cite{SageMaker}. Advanced research efforts are pushing these boundaries, advocating for more integrated and sophisticated monitoring solutions to preemptively catch errors before they escalate~\cite{wang2023gemini, wu2023transom}.

\stitle{Checkpointing.}
Checkpointing, a technique that periodically saves the training state, serves as a pivotal recovery mechanism in deep learning frameworks. The focus of recent research has been to streamline checkpointing—enhancing its efficiency to mitigate the time lost to recomputation and accelerate the recovery process in the event of a failure~\cite{paszke2019pytorch, abadi2016tensorflow, wang2023gemini, mohan2021checkfreq, eisenman2022check, nicolae2020deepfreeze}. Despite its benefits, the complexity of failures within distributed training systems often transcends what checkpointing alone can address.

\stitle{Elasticity.}
In the face of failures, some systems employ elasticity -- adjusting parallelism settings to prevent training interruptions~\cite{KungFu, Elastic, 9373916, Elan, Lyra, shukla2022singularity, jang2023oobleck, athlur2022varuna}. While this adaptability is advantageous for maintaining operational continuity, it may introduce additional overhead and potentially reduce throughput. Moreover, the complexity of integrating such elastic designs into high-performance systems like Megatron often poses significant challenges.

\stitle{Redundant Computation and Hot Spares.}
Other strategies involve employing redundant computation or allocating hot spares to preempt failures~\cite{thorpe2023bamboo, athlur2022varuna}. While these methods aim to provide a buffer against disruptions, they come with a significant economic and resource cost, highlighting the need for more efficient solutions.

\stitle{Workload Managers.}
Workload managers SLURM\cite{yoo2003slurm} and Kubernetes~\cite{bernstein2014containers} are pivotal in orchestrating general-purpose computing, offering a suite of functionalities that include cluster management, task queuing, scheduling, and efficient resource allocation. These platforms typically handle tasks as black boxes -- opaque entities with predefined static configurations -- submitting them into a queue for execution without intimate knowledge of their inner workings. While they are equipped with mechanisms like hot spares and automated retries for basic failure handling, they lack the bespoke features necessary for the specialized domain of LLM training. As a result, when failures occur, the intricate task of diagnosis and recovery largely reverts to the LLM training tasks themselves.

\begin{figure}
  \centering
  \begin{subfigure}{0.75\linewidth}
    \centering
    \includegraphics[width=0.9\linewidth]{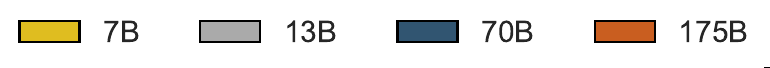}
  \end{subfigure}

  \centering
  \begin{subfigure}{0.49\linewidth}
    \centering
    \includegraphics[width=0.9\linewidth]{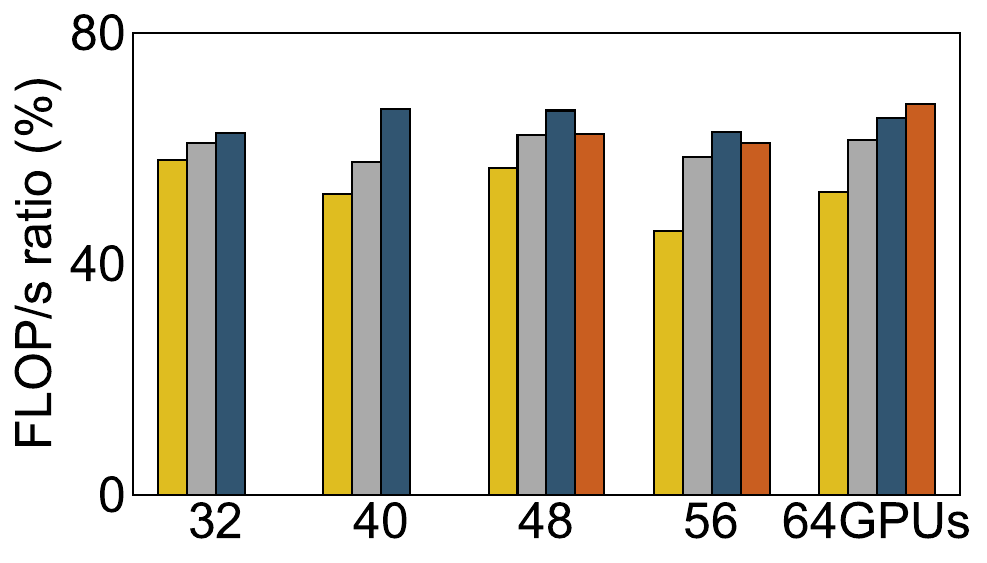}
    \label{fig:flops-utilization}
  \end{subfigure}
  \hspace*{-3ex}
  \begin{subfigure}{0.54\linewidth}
    \centering
    \includegraphics[width=0.9\linewidth]{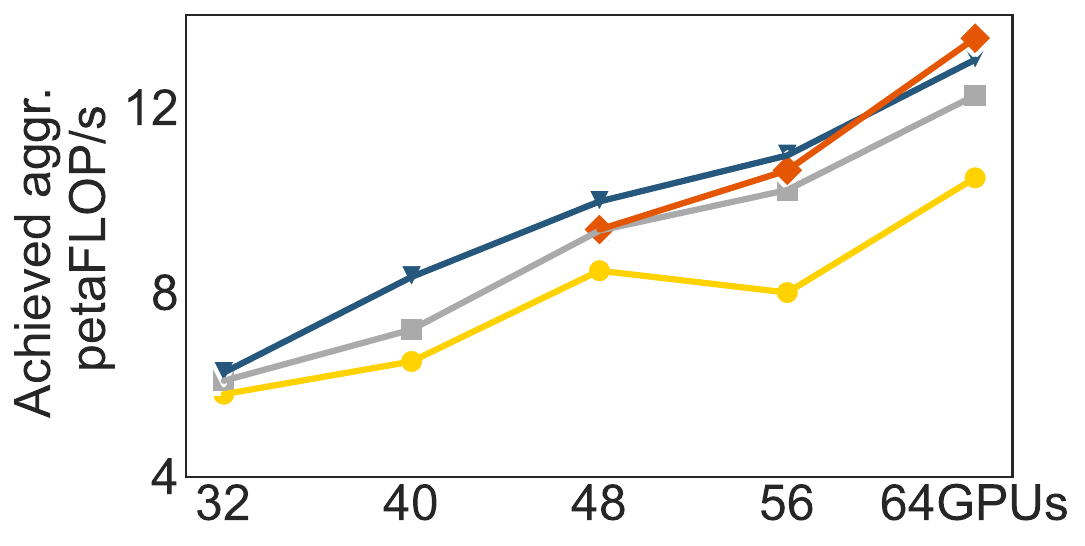}
    \label{fig:aggregated-flops}
  \end{subfigure}

  \vspace*{-1ex}
  \caption{Achieved FLOP/s ratio and aggregate FLOP/s for training varying-sized GPT-3 models using Megatron.}
  \label{fig:exp-scalability-gpus}
  \vspace*{-2ex}
\end{figure}

\subsection{Opportunities}
\label{sec:background:opportunities}
We have identified three key opportunities not yet fully realized by existing solutions, which could significantly enhance the management of LLM training failures.

\stitle{O1: Cost-Effectiveness in Failure Management.}
In LLM training, failures are not just operational setbacks; they entail excessive manual interventions and wasted training time, causing significant inconvenience as highlighted in Sections~\ref{sec:introduction} and \ref{section:background:stats}. Indeed, around 73\% of such failures can be rectified by simply restarting the training processes. Although this process takes about 68 minutes under default settings, sophisticated checkpointing methods like GEMINI\cite{wang2023gemini} offer potential reductions in downtime. Despite the mean time between failures being relatively infrequent -- ranging from a day to a week for a 128 GPU cluster -- the economic implications cannot be overlooked. Resilient training systems\cite{jang2023oobleck,athlur2022varuna,thorpe2023bamboo}, which show considerably less efficiency compared to Megatron in normal conditions (as shown in Figure~\ref{fig:training-throughput}), are not economically sustainable. Relying on such systems is akin to keeping a significant portion of the training cluster idle, which is impractical, especially in larger clusters. Additionally, the use of hot spares and redundancy, while helpful for continuity, must be balanced against their economic impact to avoid resource wastage. This situation raises a critical question: {\em how can we alleviate the pain points associated with failures while simultaneously maximizing economic efficiency and minimizing resource wastage?}

\stitle{O2: Inherent Elasticity with Non-Linear Performance.} The concept of 3D parallelism introduces a remarkable degree of flexibility and elasticity into LLM training. This framework allows for adjustments in parallelism across three distinct dimensions while preserving the integrity of the training semantics. However, this elasticity is a double-edged sword. It demands meticulous management to optimize resource utilization and enhance performance. For instance, Figure~\ref{fig:exp-scalability-gpus} illustrates the variations in achieved aggregate FLOP/s and its ratio to the theoretical peak FLOP/s of a healthy system (settings detailed in Section~\ref{sec:exp-config-plan}). A notable trend is the non-linear, and sometimes non-monotonic, relationship between the number of GPUs and the achieved aggregate FLOP/s. Adding a seemingly small number of GPUs, say 8 to a 48 GPU cluster, can lead to performance dips due to the inability to directly translate the optimal configuration from a 48 GPU setup to a 56 GPU one, mainly owing to memory constraints. Additionally, tasks operating at different scales may exhibit varying levels of resource utilization.
This phenomenon underscores a critical insight: simply maximizing the use of available resources doesn't guarantee peak performance. It also implies that altering parallelism settings can significantly impact the efficiency of each GPU in the cluster. The question then becomes: {\em how can we harness this inherent elasticity of LLM training to ensure resilient operations while simultaneously achieving high performance and efficient GPU utilization?}

\stitle{O3: Rethinking from a Workload Manager's Perspective.} Workload managers, positioned uniquely at the helm of cluster operations, can revolutionize the approach to LLM training failures. These systems, such as SLURM and Kubernetes, are not just for queueing and managing tasks; they provide a global perspective of the entire cluster, coupled with detailed node-specific information through agents. This comprehensive oversight allows for a holistic management approach, transcending the traditional method of treating training tasks as isolated, black-box entities in a queue. Such managers can seamlessly integrate with advanced training systems like Megatron, enhancing the efficacy of existing tools for checkpointing and monitoring. The real transformative potential lies in leveraging the inherent elasticity of LLM training tasks. By no longer confining these tasks to static configurations within queues, a workload manager can dynamically select optimal configurations, efficiently managing resources in response to real-time conditions. This strategy extends to handling the intricacies of failure recovery, guiding the system from detecting and addressing issues to smoothly transitioning back into Megatron's standard training process. This leads us to a crucial consideration: {\em can we innovate a workload manager that not only supports self-healing LLM training at scale but also adopts a holistic approach, maximizing economic efficiency throughout the process?}

%% file: sec-system.tex
\section{System Design}
\label{sec:system}
\begin{figure}[t!]
    \centering
    \includegraphics[width=0.8\linewidth]{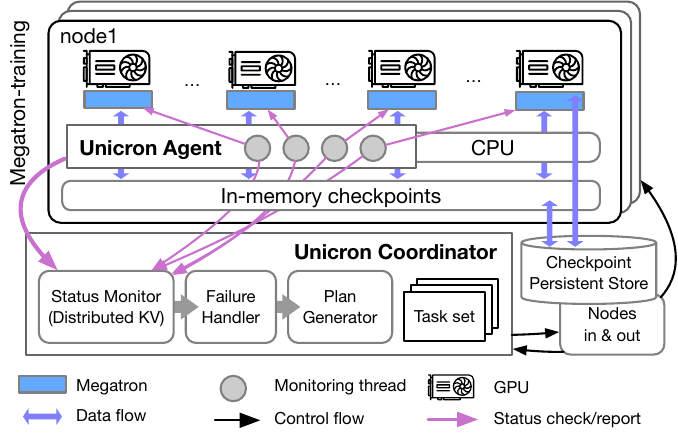}
    \caption{The system architecture of \spk.}
    \label{fig:architecture}
    \vspace*{-2ex}
\end{figure}

\stitle{Our Answer:} \spk~is designed as a distributed workload manager built with Megatron, to enhance the training process by achieving economic efficiency and self-healing capabilities.  
The architecture of \spk~is depicted in Figure~\ref{fig:architecture}.
The non-intrusive integration of \spk~with Megatron guarantees the preservation of both existing techniques and the future functionalities of Megatron, thereby ensuring efficient training task execution and maintaining strict semantics preservation.
Furthermore, \spk~introduces additional components, specifically the \emph{\spk~agent} and the \emph{\spk~coordinator}, as well as several key techniques to enable efficient self-healing in the event of various failures during LLM training.
Additionally, \spk~considers concurrent training tasks within the cluster, with the goal of improving overall resource utilization.
The combination of these features ensures that \spk~maintains economic efficiency while effectively achieving self-healing capabilities.
Next, we will discuss the key components and techniques employed by \spk~in detail.

\subsection{\spk~Agent}
The \spk~agent is a crucial component, performing several key tasks associated with a single machine in the cluster.

\stitle{Error Detection.} The \spk~agent establishes a persistent connection with the coordinator, which helps to ensure the continuous availability and responsiveness of the node. 
Besides, it assigns a dedicated monitoring thread on the CPU for each GPU. These monitoring threads closely observe the status of the GPU, capturing important events and detecting any irregularities or exceptions. 
This fine-grained monitoring helps in identifying potential failures or issues promptly.

\stitle{Recovery Actions Execution.} The \spk~agent plays a significant role in executing recovery actions inside a machine, guided by the \spk~coordinator and following a transition strategy. 
These actions aim to swiftly transition the system from its current state to a new state, ensuring the system is rapidly restored to a functional state.

\stitle{Checkpointing.} Recognizing the necessity of checkpointing for failure recovery as a bottom-up solution, \spk~incorporates a checkpointing mechanism. 
The \spk~agent manages the checkpointing workflow, handling in-memory checkpoints following the methodology proposed by GEMINI~\cite{wang2023gemini}, 
and asynchronously transfers these checkpoints to a remote persistent storage.
This hierarchical checkpointing approach guarantees the ability to restore training progress in the event of failures. 
It is noteworthy that our framework operates independently of research aimed specifically at optimizing the checkpointing process.
Advanced mechanisms can be integrated into \spk~to reduce the overhead associated with checkpointing, as our focus lies elsewhere.

\subsection{\spk~Coordinator}
\label{sec:scheduler}
The \spk~coordinator leverages the information collected by the agents to make informed decisions and coordinate actions across the entire cluster. Its responsibilities include:

\stitle{Consolidation of Process Status.} The \spk~coordinator utilizes a distributed key-value store (implemented using etcd~\cite{etcd}) referred to as the status monitor, to consolidate and store the process statuses reported by the monitoring threads from each agent, allowing for a comprehensive view of the system's health.

\stitle{Error Handling.} Upon detecting abnormal status from the status monitor, the \spk~coordinator assesses the situation to determine the correct response. It then directs the agent to implement the necessary actions, ensuring that these actions are well-coordinated and synchronously executed across the workers involved.

\stitle{Reconfiguration Plan Generation.} The \spk~coordinator plays a crucial role in developing an optimal reconfiguration plan when necessary, such as in cases of node faults, node integration, task finished, or task launched within the cluster. This plan is guided by a comprehensive model that considers all ongoing tasks within the cluster, with the primary objective of minimizing any potential decline in training efficiency.

\stitle{Training Task Management.} The \spk~coordinator is responsible for managing the training tasks within the cluster. It keeps a watch on the status of each task, coordinates task submission and termination with the cloud service, utilizing a task set for tracking.
Each task is associated with the minimum computational requirements, as well as a weight assigned by the user to model its priority. The weights can represent task priority or the unit price / quota the users willing to pay for each unit of useful FLOP/s. These weights and computational requirements serve as essential inputs for the coordinator to determine the optimal reconfiguration plan for distributing available resources among the tasks.

\stitle{Other External Interactions.} In addition to its primary responsibilities, the \spk~coordinator also handles other external interactions. These interactions encompass a range of activities such as alerting maintenance personnel of failure incidents, acknowledging node recovery updates, incorporating newly provisioned nodes from the cloud service, and others.

\subsection{Key Techniques}
\spk~incorporates several key techniques to address the three types of costs related to failure recovery. 
Firstly, the cost of detecting errors, denoted as $C_{\text{detection}}$, is managed by the error detection module described in Section~\ref{sec:detect}. This module leverages four distinct detection methods to ensure failures are identified promptly and accurately, while avoiding extra overhead on the training process.
Next, the actions taken to respond to failures, including reattempting in-place, restarting the training process, and reconfiguring the cluster, contribute to the transition cost, $C_{\text{transition}}$. 
\spk~seeks to minimize this cost through the adoption of a rapid transition strategy, which is explained in Section~\ref{sec:transition}.
Lastly, the cost of sub-healthy, referred to as $C_{\text{sub-healthy}}$, is associated with reduced resource utilization and training efficiency due to sub-optimal configurations.
This cost can have a prolonged impact on overall system performance. To mitigate it, \spk~utilizes a formulated model designed to determine the most effective configuration plan. The development and solution of this model are discussed in detail in Section~\ref{sec:model}.

%% file: sec-detection.tex
\section{Error Detection and Handling}
\label{sec:detection}
In this section, we present the design for error detection and handling strategies in \spk.
Effective error detection methods are essential for minimizing detection costs, while tailored handling strategies ensure appropriate measures for failure recovery. 
Table~\ref{tab:classification} summarizes the four detection methods employed by \spk, along with our subjective classification of error statuses based on their severity levels.
The severity levels represent the impact of the error on the training process, ranging from \emph{\sev{1}} (most severe) to \emph{\sev{3}} (least severe), and are used to determine the appropriate handling strategy.

\begin{table}[h]
  \centering
  \caption{Detection methods and severity levels of errors.}
  \label{tab:classification}
  \fontsize{8}{9}\selectfont
  \begin{tabular}{|p{2.8cm}|l|l|}
    \hline
    \textbf{Detection method}   & \textbf{Error status} & \textbf{Severity} \\
    \hline
    Node health monitoring      & Lost connection & \sev{1} \\
    \hline
    Process supervision         & Exited abnormally & \sev{2} \\
    \hline
    \multirow{9}{2.8cm}{Exception propagation}
                               & Connection refused/reset & \sev{3} \\
                               & Illegal memory access & \sev{2} \\
                               & ECC errors & \sev{1} \\
                               & Invalid DMA mapping & \sev{1} \\
                               & CUDA errors & \sev{2} \\
                               & NVLink errors & \sev{1} \\
                               & GPU driver errors & \sev{1} \\
                               & Other network errors & \sev{3} \\
                               & Other software errors & \sev{2} \\
    \hline
    \multirow{4}{2.8cm}{Online statistical monitoring}
                               & NCCL timeout & \sev{3} \\
                               & Link flapping & \sev{3} \\
                               & Task hang & \sev{2} \\
                               & Other software errors	& \sev{2} \\
    \hline
  \end{tabular}
\end{table}

\subsection{Error Detection}
\label{sec:detect}
\spk~utilizes in-band error detection by continuously monitoring the real-time status of each training processes. 
This is done through the monitoring threads of the agent, which track training progress, exceptions and communication timeouts.
The agent is launched at the beginning of the training process and operates concurrently with it.
It enables the system to promptly identify any irregularities or exceptions. 
Compared to other solutions that rely on out-of-band monitoring like cloud monitoring services, this method provides a significant advantage in terms of efficiency and accuracy.
Furthermore, the monitoring threads operate on the CPU, ensuring that they introduce no extra load on the GPU, which carries out the primary training workload.

\stitle{Node Health Monitoring.} The persistent connection maintained between the \spk~agent and the \spk~coordinator guarantees node availability. If this connection is lost, the node is marked as unavailable, and a \sev{1} failure is triggered.

\stitle{Process Supervision.} The \spk~agent has a monitoring thread for each GPU to watch the training process. Should a process terminate unexpectedly, this thread signals a \sev{2} failure to the coordinator for resolution.

\stitle{Exception Propagation.} Exception handling is crucial for the prompt detection of incidental errors, such as ECC errors, NVLink errors, CUDA errors, and others. These errors are immediately identified once the GPU issues an exception, which is then captured by the monitoring thread and reported to the coordinator.

\stitle{Online Statistical Monitoring.} \spk~leverages online statistical monitoring to detect certain errors like NCCL timeout, TCP timeout, and task hangs, where notifications are delayed. 
For instance, a delay of up to $30$ minutes may occur before a NCCL timeout error is raised, as shown in Figure~\ref{fig:stage}. 
Although the timeout threshold can be adjusted, it is challenging to determine the appropriate value, as it is highly dependent on the system configuration and the workload.

\begin{figure}[t]
  \centering
  \includegraphics[width=0.9\linewidth]{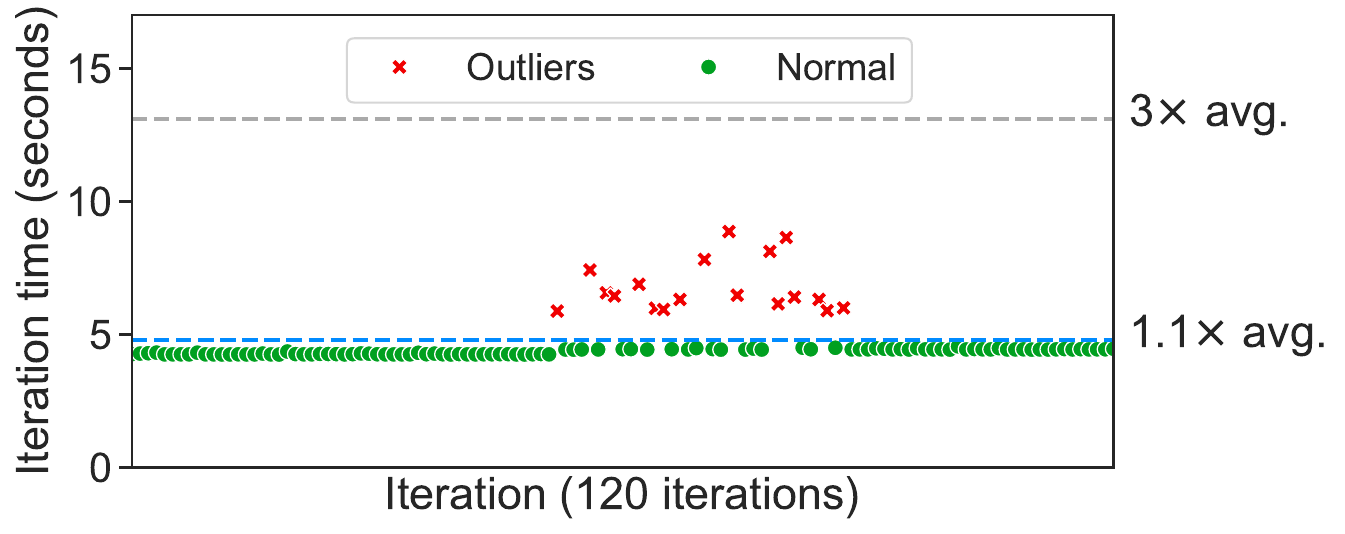}
  \caption{Completion time per iteration.}
  \label{fig:detect}
  \vspace*{-3ex}
\end{figure}

The monitoring threads implement online statistical monitoring to detect these errors.
Under normal training conditions, which proceed periodically, the statistics of iterations should reveal a relative consistency when the system is set to a particular configuration, as Figure~\ref{fig:detect} shows with green dots representing iteration completion times for training a GPT-3 $175B$ model on $256$ NVIDIA H800 GPUs.
While minor fluctuations may occur due to network variations and congestion, they typically stay within a reasonable margin indicated by the blue line (i.e., $1.1\times$ the average iteration time).
The red dots illustrate instances where a network switch has been deliberately turned off, leading to a marked increase in completion time; yet, the training process manages to persist.
If the waiting time surpasses the threshold denoted by the grey line, this confirms a failure, requiring immediate recovery measures. 
Empirical evidence suggests that setting the failure threshold at $3\times$ the average iteration time, achieves a practical balance between efficiency and accuracy.

\subsection{Error Handling}
\label{sec:self-healing}

\begin{figure}[t]
    \centering
    \includegraphics[width=\linewidth]{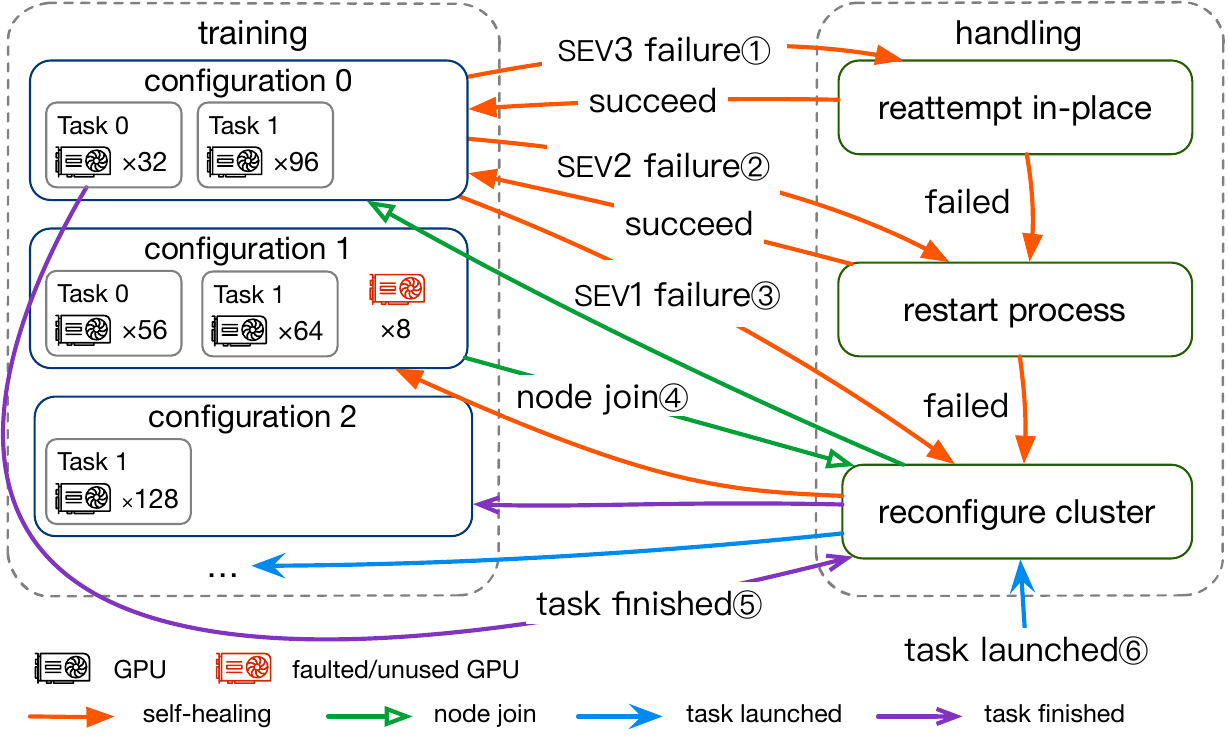}
    \caption{The error handling workflow in \spk.}
    \label{fig:workflow}
    \vspace*{-2.5ex}
\end{figure}

When any abnormal status is detected, the \spk~coordinator is notified and proceeds to take  appropriate actions.
The first step in handling a failure is to classify the collected status based on its severity level, as outlined in Table~\ref{tab:classification}.
This classification (indicated by \ding{172} to \ding{174}) helps determine the appropriate recovery strategy, which corresponds to one of three specific actions, following the guidance of Figure~\ref{fig:workflow}.
Additionally, Figure~\ref{fig:workflow} includes other triggers (indicated by \ding{175} to \ding{177}), to ensure the smooth operation of the training process.

\stitle{Reattempt In-place.} The initial attempt to mitigate a failure involves retrying the operation where it failed, assuming there are no indications of underlying software or hardware issues, which is classified as \sev{3} failure \ding{172}. 
This method is effective for addressing issues like temporary connection problems (e.g., link flapping or connection refused/reset). 
If the reattempt succeeds, the training process immediately proceeds as normal. 
Otherwise, the issue is upgraded to a \sev{2} failure. 

\stitle{Restart Process.} \sev{2} failures \ding{173}, such as CUDA errors or illegal memory accesses, are resolved by restarting the training process on the affected node. 
The system configuration, including the number of GPUs, parallelism settings, and process ranks, remains unchanged.
And the training states are recovered from either other data parallel replicas or the latest checkpoint, as detailed in Section~\ref{sec:transition}.
If the restart attempt fails, the severity of the failure is upgraded to \emph{\sev{1}}.

\stitle{Reconfigure Cluster.} In the event of a \sev{1} failure \ding{174}, the \spk~coordinator isolates the failed node and initiates a reconfiguration procedure.
This happens when the coordinator receives notifications of GPU node faults or if the system fails to recover from a \emph{\sev{2}} failure. 
When a previously failed and unused node is successfully recovered and becomes available again, or when a new node is allocated, it can \emph{join \ding{175}} the ongoing training process.
In such cases, the \spk~coordinator initiates the cluster to enter the reconfiguration process, which allows the system to adapt and incorporate the additional available resources.
Additionally, the reconfiguration process can also be triggered when an existing task is \emph{finished \ding{176}} or a new task is \emph{launched \ding{177}}.
This is because the optimal configuration plan may differ from the current configuration, requiring the reallocation of resources to achieve the global optimal training efficiency of the cluster.

%% file: sec-model.tex
\newcommand{\tDetectFailure}{T_{\text{detect\_failure}}}
\newcommand{\tScaleOut}{T_{\text{scale\_out}}}
\newcommand{\tScaleIn}{T_{\text{scale\_in}}}
\newcommand{\tWait}{T_{\text{wait\_for\_available\_resources}}}
\newcommand{\tInit}{T_{\text{initialize\_new\_worker}}}
\newcommand{\tHtoSub}{T_{\text{healthy}\rightarrow\text{subhealthy}}}
\newcommand{\tSubhealthy}{T_{\text{subhealthy\_duration}}}
\newcommand{\tSubtoH}{T_{\text{subhealthy}\rightarrow\text{healthy}}}
\newcommand{\tCorruption}{T_{\text{corruption}}}

\newcommand{\Throughput}{T}
\newcommand{\Benefit}{F}
\newcommand{\Cost}{C}
\newcommand{\Time}{D}
\newcommand{\Task}{t}

\section{Optimal Reconfiguration Plan Generation}
\label{sec:model}
This section introduces a method for generating an optimal reconfiguration plan to efficiently distribute tasks across GPUs within a distributed training cluster. The formulation of this plan is critical as it directly influences the sub-healthy cost $\Cost_{\text{sub-healthy}}$. To capture these costs, we define the metric of \benefit, representing the weighted achieved aggregate FLOP/s, and formulate an optimization problem to minimize the costs incurred during sub-healthy and transition states.

\subsection{Model Formulation}
With $n$ workers available in the cluster for distributed training and $m$ tasks to be trained, 
our goal is to fully utilize the computation capacity of the resources while meeting the requirement of each running task. 
A GPU card is considered as a worker by default within the scope of this problem. 
Before delving into our model, we first define the metric to measure the training efficiency of a task. 

\stitle{\benefit.} 
The \benefit~of a task measures the \underline{w}eighted achieved \underline{a}ggregate \underline{F}LOP per second.
Formally, we define a function $F:\mathbb{N}\times \mathbb{N}\rightarrow\mathbb{R}$, where $F(t,x)$ represents the \benefit~of task $t$ when $x$ workers assigned to it.

As implied by the name, the most important part of \benefit~is the \emph{achieved aggregate FLOP/s}, denoted as $T(t,x)$, for a given task $t$ with $x$ workers. 
Notice the $T(t, x)$ reflects the optimal performance of the task, which is obtained by tuning the complex combination of parallelism hyperparameters and other optimization settings. 
To address this, we rely on calibrating tasks on the given GPU cluster and leverage automatic execution plan generation techniques~\cite{zheng2022alpa} to estimate the optimal parallelism settings and associated $T(t, x)$.

In addition to the achieved aggregate FLOP/s, we further integrate two additional factors to model the minimum computational requirements and task priorities.
\emph{Requirement condition} ($T_{\text{necessary}}(t)$) represents the minimum required resources for a task given. A task will only be scheduled if the available resources can satisfy this requirement.
\emph{Task weight} ($w(t)$) is used to model the priority of a task. Tasks with higher priority are assigned higher weights. By default, we set $w(t)=1$ for all tasks. This weight factor allows the user to adjust the relative importance of different tasks in the problem, with the recommended values for $w(t)$ range between $0.5$ and $2.0$.

Accordingly, the \benefit~is defined as follows:
\begin{equation}
  \label{eq:benefit}
  F(t,x) = \left\{
    \begin{aligned}
      &w(t) \cdot T(t,x) &&\text{if } (t,x) \vdash T_{\text{necessary}}(t), \\
      &0 &&\text{otherwise}.
    \end{aligned}
  \right.
\end{equation}
Here, $F(t,x)$ is calculated as the product of the task's weight, $w(t)$, and its achieved aggregate FLOP/s, $T(t,x)$, when $(t,x)$ satisfies the necessary requirements, $T_{\text{necessary}}(t)$.
Otherwise, if it falls short of this threshold, the \benefit~of the task is considered to be zero.

\stitle{Optimization Objective.}
The formulation of the optimization problem for plan generation is centered around a trade-off: maximizing the \benefit~of the cluster post-reconfiguration while minimizing the impact on GPUs during the transition. 

\begin{equation}
  \label{eq:opt-simplified}
  \begin{aligned}
    \mathop{\arg\max}_{x_1',\ldots,x_m'}\quad 
    & \sum_{i=1}^m G(t_i,x_i'), \\
    \text{where}\quad G(t_i,x_i') =& F(t_i,x_i') \cdot D_{\text{running}}(n') \\
    -& F(t_i,x_i) \cdot \mathbf{1}(t_i,x_i \rightarrow x_i') \cdot D_{\text{transition}}, \\
    \text{subject to}\quad & \sum_{i=1}^m x_i' \leq n'.
  \end{aligned}
\end{equation}

Prime notation (') differentiates states before and after reconfiguration. The subscript \(i\) denotes the task identifier, \(t_i\) represents the \(i\)-th task, and \(x_i\) the number of workers initially assigned. The constraint ensures that the workers assigned across all tasks do not exceed the total available in the cluster. The goal is to maximize the cluster's cumulative reward, represented by the sum of individual task rewards \(G(t_i,x_i')\).

The primary term, \(G(t_i,x_i')\), reflects the reward when the cluster is operational post-reconfiguration, calculated as the \benefit~of task \(t_i\) with \(x_i'\) workers, multiplied by the expected run duration \(D_{\text{running}}(n')\). This duration is contingent on the operational condition of GPUs and the cluster size; a larger pool of GPUs implies a higher likelihood of failure, potentially shortening the run duration.

The penalty term, \(F(t_i,x_i) \cdot \mathbf{1}(t_i,x_i \rightarrow x_i') \cdot D_{\text{transition}}\), captures the \benefit~loss during transition, where \(D_{\text{transition}}\) is the estimated duration of this period. The indicator function \(\mathbf{1}(t_i,x_i \rightarrow x_i')\) activates when there is a change in the number of workers assigned to task \(t_i\) or if a worker fault occurs:

\begin{equation}
  \label{eq:indicator}
  \mathbf{1}(t_i,x_i\rightarrow x_i') = \left\{
    \begin{aligned}
      &1 && \text{if } x_i \neq x_i' \text{ or a worker of \(t_i\) faults}, \\
      &0 && \text{otherwise}.
    \end{aligned}
  \right.
\end{equation}

This term discourages frequent reconfiguration, especially for healthy tasks, by factoring in the \benefit~that could have been achieved by unaffected GPUs. 

The significance of each term is context-dependent. In stable clusters with fewer faults, maximizing \benefit~takes precedence, focusing on the reward during healthy operation. Conversely, in larger or less reliable clusters where faults are more common, the penalty term becomes more critical, emphasizing the need to limit the scope of reconfigurations and maintain as many tasks running as possible.

\subsection{Solving Algorithm}
\label{sec:solver}
This problem can be solved through the dynamic programming algorithm, in which we define the state as $S(i,j)$, representing the maximum value of the first $i$ tasks with $j$ workers.
Accordingly, the state transition equation is given by:
\begin{equation}
  \label{eq:dp}
  S(i,j) = \max_{k=0}^{j}\{S(i-1,j-k)+G(t_i,k)\}
\end{equation}

This equation implies that the maximal reward of the first $i$ tasks with $j$ workers can be transited from the maximal reward of the first $i-1$ tasks with $j-k$ workers plus $G(t_i,k)$, where $k$ denotes the number of workers assigned to the $i$-th task.
We set $S(0,j)=0$ for $j>0$, to initialize the states.
The optimal value of Equation~\ref{eq:opt-simplified} can be obtained from $S(m,n')$.
To obtain the optimal assignment strategy specifically, we can trace back the state transition process.

\stitle{Complexity.} The time complexity of this algorithm is $O(mn^2)$, where $m$ represents the number of tasks and $n$ represents the cluster size.
In practice, both $m$ and $n$ remain moderate in size, making this algorithm efficient to execute.
Furthermore, the \spk~coordinator can pre-compute and prepare the lookup table for the dynamic programming algorithm in advance, taking into account potential failure scenarios of any task or joining node. This allows for one-step advancement from the current configuration.
Once reconfiguration is required, the \spk~coordinator can directly retrieve the optimal assignment strategy from the lookup table, thereby reducing the time complexity to $O(1)$.

%% file: sec-transition.tex
\section{Transition Strategy}
\label{sec:transition}
In this section, we delve into the implementation of transitioning a training task to a new configuration. We first review the training process of an iteration in Megatron to identify the maximal possible partial results that can be reused. Next, we present the adaptive resumption process according to failure scenarios. This resumption can leverage the partial results to finish the current iteration.
Finally, we discuss the transition process for a task that is instructed to reconfigure. In this process, we aim to minimize the state migration cost by leveraging the nearest available states.

\subsection{Iterations in Megatron}
\label{subsec:adaptive}
\begin{figure}[t]
  \centering
  \includegraphics[width=0.9\linewidth]{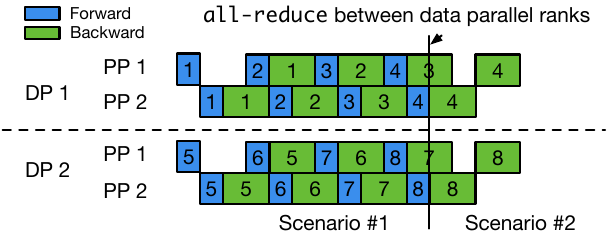}
  \caption{Timeline of data and pipeline parallelism.}
  \label{fig:pipeline-data-parallelism}
\end{figure}

We first review the training process of an iteration in Megatron, which is depicted in Figure~\ref{fig:pipeline-data-parallelism}.
In an iteration, a \emph{global-batch} of samples will be forward passed through the model, and the gradients will be accumulated until the completion of the backward pass.
For distributed training, a global-batch is further partitioned into multiple \emph{micro-batches}, e.g., 8 micro-batches in the figure.

Corresponding to the parallelism of the model, a rank in distributed data parallelism (DP) is responsible for a subset of the micro-batches.
With the DP degree of $DP$, and $B$ micro-batches in a global-batch, each DP rank is responsible for $k = B / DP$ micro-batches.
Therefore, we utilize two dimension of indexes to denote the $j$-th micro-batch in the $i$-th DP rank, i.e., $grad_{i,j}$.
The aggregated gradient for the global-batch, denoted as $grad$, is computed as follows:

\begin{equation}
\label{eq:grad}
  grad = \underbrace{\sum_{i=1}^{DP}}_{\text{all-reduce}} \underbrace{\sum_{j=1}^{k} grad_{i,j}}_{\text{accumulation}}
\end{equation}

Within a rank of DP, a rank of PP handles a subset of the layers in the model, a.k.a., stage. 
As shown in the figure, the micro-batch 1 is distributed to the DP 1, it first forward passes through the PP 1 then the PP 2, and the backward pass reverses the order. 
The PP pipeline processes the micro-batches within a DP rank and accumulate the gradients until the completion of the backward pass for the last micro-batch. 
Subsequently, the parameters are updated by averaging the gradients from all DP ranks through the \texttt{all-reduce} operation. 
Notice, that the tensor parallelism (TP) can be further employed within each PP rank, which is not shown in the figure as it does not impact our strategy.

\subsection{Resuming from a Failed Iteration}
\label{subsec:resume}
Failures can disrupt training at any point, but leveraging partial results from completed micro-batches within the current global-batch can mitigate recomputation costs.
Thanks to the gradient accumulation in micro-batch iterations in the distributed data parallelism, as shown in Equation~\ref{eq:grad}, a state-driven approach has been devised to resume training from a failed global-batch iteration.

Failures may occur in two scenarios: prior to the \texttt{all-reduce} operation (scenario \#1) or subsequent to it (scenario \#2), as depicted in Figure~\ref{fig:pipeline-data-parallelism}.
\spk~uses a micro-batch iteration scheduler that monitors and synchronizes progress, enabling the training process resume seamlessly from the point of failure. 
It also redistributes micro-batches from the failed DP rank to others in a round-robin fashion, ensuring the integrity of the training process.

\stitle{Scenario \#1.} Before initiating the \texttt{all-reduce}, each DP rank manages its own accumulated gradients and independently performs forward and backward passes for micro-batches assigned to it.
\spk~tracks each DP rank's progress through forward and backward passes. 
In the event of a failure, the resumption process consists of the following steps: 1) pausing the training process, 2) re-establishing the network connection among healthy workers, 3) redistributing the micro-batches owned by the failed DP rank, and 4) resuming the training process.
After redistributing, each remaining DP rank now owns $k' = k+ k / (DP - 1)$ micro-batches. Consequently, the aggregated gradient for the global-batch is computed as follows:
\begin{equation}
  \label{eq:grad2}
    grad = \sum_{\substack{i=1 \\ i \neq x}}^{x-1}\bigg( \underbrace{\sum_{j=1}^{k} grad_{i,j}}_{current} + \underbrace{\sum_{j=k+1}^{k'} grad_{i,j}}_{redistributed} \bigg)
\end{equation}

\stitle{Scenario \#2.} In the event of a failure occurring after the \texttt{all-reduce} operation started, the response depends on whether the failed worker's gradients have been reduced, due to pipeline parallelism inside each DP rank. 
1) If the worker's gradients were already reduced, the worker can be omitted, as the aggregated gradient is already accounted for by the other DP replicas. Training proceeds uninterrupted.
2) Conversely, if the gradients from the failed worker were not yet reduced, \spk~ensures semantic correctness by redistributing the failed DP rank's micro-batches to remaining DP ranks, similar to scenario \#1, and recomputing the aggregated gradient. 
Unfortunately, different from scenario \#1 where no gradient has been reduced, there are partial gradients (segmented on the stage/layer) that have already been reduced and we need to distinguish them from the rest unreduced gradients. Accordingly, when recomputing the redistributed micro-batches, we only need to recompute the unreduced gradients and ensure the reduced gradients are not overwritten. 
It's worth noting that the \texttt{all-reduce} operations solely take place at the end of a global-batch iteration and occupy a minor fraction ($<2\%$ when training the GPT-3 175B model using 128 GPUs) of the iteration time. Therefore, the likelihood of a failure occurring after the initiation of the \texttt{all-reduce} operation remains relatively low.

\subsection{Transitioning to the New Configuration}
Once the parameter updates for the ongoing iteration are completed, or the problematic nodes have been fixed and reintegrated into the cluster. \spk~instructs the workers to transition to the new configuration planned by the plan generator. Next, we investigate the major problem in transitioning a given task: how to migrate the training states to the new configuration.

According to Equation~\ref{eq:indicator}, there are two kinds of tasks requiring transitioning: 1) a worker of the task is faulted, and 2) the task is instructed to scale in or out.
Regarding the first case, \spk~follows the nearest principle to minimize the state migration cost.
\spk~initially attempts to request the state of the healthy rank of DP, since the state is already replicated in each DP rank.
If replication cannot be fulfilled, \spk~resorts to loading the necessary data from the hierarchical checkpoints~\cite{wang2023gemini}.
Experience from GEMINI suggests that it is highly likely that the in-memory checkpoint is available, avoiding slower access from remote storage and enabling quick resumption of training.
In contrast, if the involved task is the one that runs normally, complete parameters and optimizer states must already be present in the cluster. \spk~then requests the workers to proactively replicate the state from existing workers.
Notice, that different workers issue replication requests simultaneously to minimize their own transition period.

%% file: sec-evaluation.tex
\section{Evaluation}
\label{sec:evaluation}
In this section, we evaluate \spk's performance using a range of workloads and failure scenarios. We first conduct micro-benchmarks to measure \spk's error detection time, transition time, the effective throughput and the \benefit~metric.
These tests showcase how effectively \spk's techniques tackle failure recovery challenges. 
Additionally, we compare the overall training efficiency of \spk's with that of established baselines under diverse failure traces, in a multi-task environment over a specified period.

\begin{table}[h]
  \centering
  \caption{The time to detect different kinds of failures.}
  \label{tab:exp-detection}
  \setlength{\tabcolsep}{5pt}
  \fontsize{8}{9}\selectfont
  \begin{tabular}{|c|l|l|l|}
      \hline
      \textbf{Case} & \textbf{Method}               & \textbf{\spk}                     & \textbf{w/o \spk}         \\
      \hline
      1             & Node health monitoring        & $5.6$ seconds                     & $5.7$ seconds             \\
      2             & Process supervision           & $1.8$ seconds                     & $D_\text{timeout}$\tnote{b}     \\
      3             & Exception propagation         & $0.3$ seconds                     & $D_\text{timeout}$              \\
      4             & Online statistical monitoring & $3\times D_\text{iter}$\tnote{a}  & $D_\text{timeout}$              \\
      \hline
  \end{tabular}
  \begin{tablenotes}
    \item[a] $D_\text{iter}$: the average time of one training iteration, typically within $1$ minute.
    \item[b] $D_\text{timeout}$: the timeout threshold of Megatron, $30$ minutes by default.
  \end{tablenotes}
  \vspace*{-3ex}
\end{table}

\subsection{Experimental Setup}
\stitle{Platform.} All experiments are conducted on the \alibaba~platform.
For the evaluation, we utilize a total of $16$ instances, with each instance equipped with 8 NVIDIA A800 (80GB) GPUs, 96 CPU cores, and 1,600 GB of CPU memory. 
The 8 GPUs inside each instance are interconnected through NVSwitch, and each instance is connected to others via four 200Gbps Ethernet NICs. 
We use \alibaba's cloud filesystem service as the remote persistent storage for checkpointing, which supports a maximum throughput of $20$ GB/s.
For the implementation of \spk, the software versions used is Megatron v23.08, PyTorch v1.12.1 and CUDA v11.3.

\stitle{Workloads.} In our evaluation, we utilize the GPT-3 model~\cite{openai2020gpt3} as the primary workload, due to its status as one of the largest and most popular models in the LLM domain. 
To represent different scales, we vary the size of the model parameters, specifically considering model sizes of $1.3$ billion, $7$ billion, $13$ billion, $70$ billion, and $175$ billion parameters.

\stitle{Baselines.} For comparison, we have selected several representative systems as baselines.
These baselines are chosen based on their availability and their ability on failure recovery.
The first is Megatron (v23.08) without any additional optimizations introduced by \spk. This baseline represents the solution of terminating the training process and restarting from the last persistent checkpoint when resources are recovered, without support for training with reduced resources.
The second baseline is Oobleck~\cite{jang2023oobleck}, which is a state-of-the-art framework that adopts dynamic reconfiguration to enable resilient distributed training with guaranteed fault tolerance.
Varuna~\cite{athlur2022varuna} is included as another baseline, which enables asynchronous checkpointing and dynamic reconfiguration based on job morphing for fast recovery.
Additionally, we also evaluate Bamboo~\cite{thorpe2023bamboo}, as the state-of-the-art solution for fault-tolerant distributed training through redundant computation.
For Oobleck, Varuna and Bamboo, the latest open-source versions provided by the respective authors have been utilized.

\begin{figure}[t]
  \centering
  \includegraphics[width=0.8\linewidth]{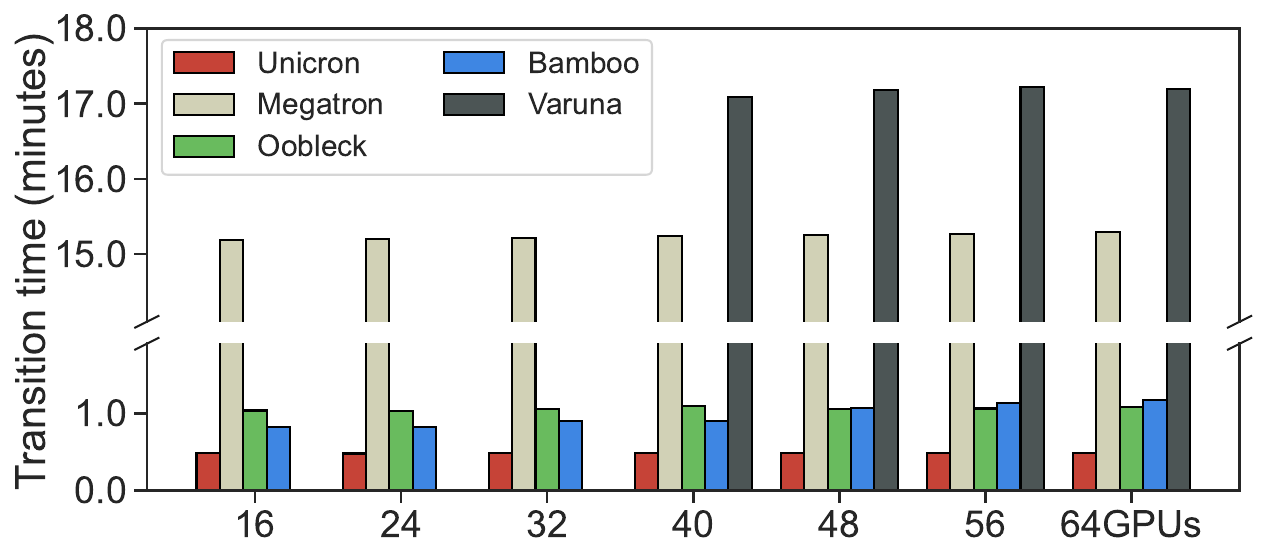}
  \caption{The transition time under failures.}
  \label{fig:exp-transition}
\end{figure}

\begin{figure*}[t]
  \centering

  \begin{subfigure}{.59\linewidth}
    \centering
    \includegraphics[width=0.3\linewidth]{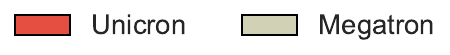}
  \end{subfigure}
  \begin{subfigure}{.4\linewidth}
    \includegraphics[width=.88\linewidth,right]{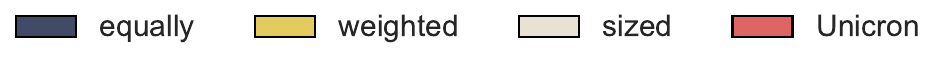}
  \end{subfigure}

  \begin{subfigure}{.36\linewidth}
    \centering
    \includegraphics[width=\linewidth]{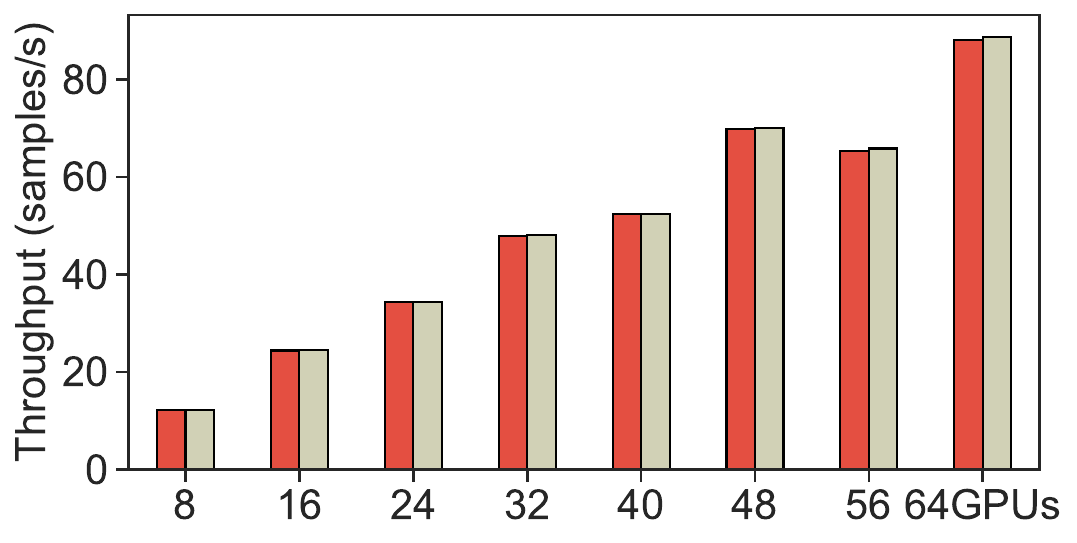}
    \caption{Training throughput (single task).}
    \label{fig:exp-throughput}
  \end{subfigure}
  \hspace*{1ex}
  \begin{subfigure}{.22\linewidth}
    \centering
    \includegraphics[width=\linewidth]{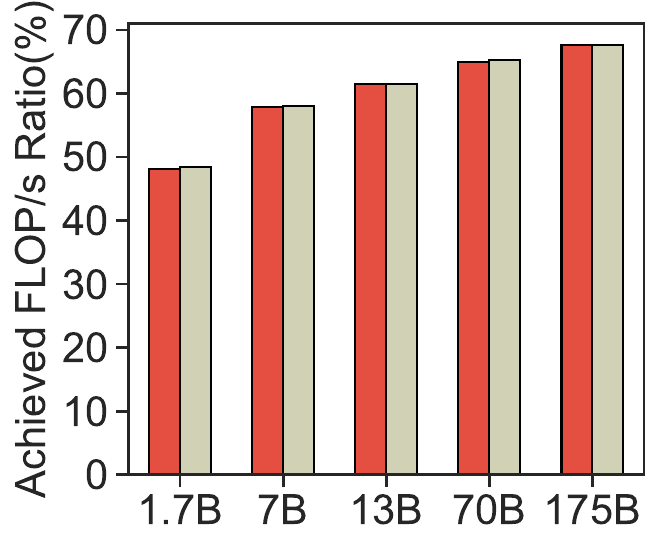}
    \caption{FLOP/s ratio (single task).}
    \label{fig:exp-outback-overhead}
  \end{subfigure}
  \begin{subfigure}{.36\linewidth}
    \centering
    \includegraphics[width=\linewidth]{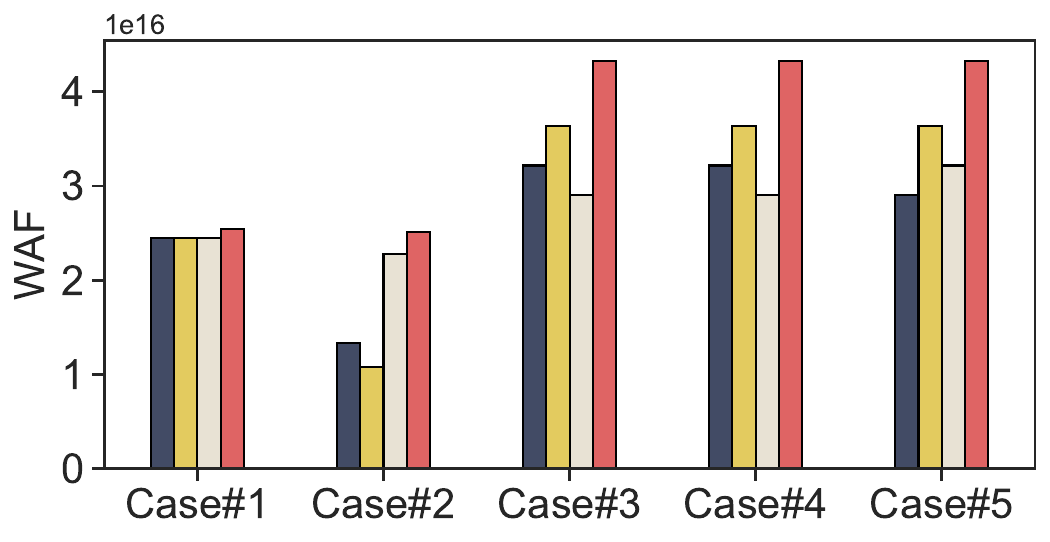}
    \caption{\benefit~(multiple tasks).}
    \label{fig:exp-benefits}
  \end{subfigure}

  \caption{Training throughput, achieved FLOP/s ratio and \benefit~of \spk~and baselines under different workloads.}
  \label{fig:exp-training-efficiency}
  \vspace*{-1ex}
\end{figure*}

\subsection{Error Detection Efficiency}
To evaluate the error detection efficiency of \spk, we simulate four failure cases, by killing a node (case 1), killing a process (case 2), throwing an exception (case 3) during training, and triggering a performance degradation (case 4).
These cases cover the most common types of failures that can occur during training, and are detected by the four detection methods implemented in \spk.
The experiments are performed using different tasks and cluster sizes, and the detection time remains relatively consistent for each case, except for case 4.
In case 4, the detection time was influenced by the average duration of one training iteration ($D_\text{iter}$).

We compare the detection time of \spk~with the baseline approach, which is Megatron without integrating \spk.
For the baseline approach, the detection time is measured as the duration from the occurrence of the failure to the termination of the training task. 
The results in Table~\ref{tab:exp-detection} demonstrate that the detection time of \spk~aligns with the baseline approach for case 1, while is significantly shorter for the remaining three cases.

\subsection{Transition Efficiency}
\label{sec:exp-transition}

Figure~\ref{fig:exp-transition} presents the transition time of \spk~compared to the baselines when a \sev{1} failure is detected during the training of the GPT-3 $7B$ model on clusters of varying scales~\footnote{As Megatron lacks support for dynamic reconfiguration, our testing of Megatron includes the use of a hot spare node that substitutes for the failed node. Consequently, the time Megatron spends waiting for resources is not factored into the transition time measurement.}.
Missing bars indicate that the training task cannot be successfully launched for certain baselines at the specified cluster size.
The transition time refers to the duration from detecting the failure to resuming training, which includes recovering the training task based on a new configuration (with a reduced number of nodes) and recomputing lost progress.

For Megatron and Varuna, the transition time required is considerably long because the training task needs to be restarted from the last checkpoint, and the progress since the last checkpoint must be recomputed~\footnote{
  Average recomputation time is 15$mins$ for 30$mins$ checkpoint intervals.}  
Oobleck and Bamboo can reduce the transition time by enabling dynamic reconfiguration, which eliminates the need to load the checkpoint and restart from it. 
However, their transition time still exceeds that of \spk.
\spk~further decreases the transition time by utilizing the transition technique, which maximizes the reuse of partial results from ongoing training iterations, thereby minimizing the loss caused by failures.
Besides, \spk~is able to maintain a relatively stable transition time across different cluster sizes by efficiently migrating training states between nodes, leveraging the characteristics of the parallelism approaches.

\subsection{Training Throughput and \benefit}
\label{sec:exp-config-plan}

In this subsection, we evaluate the training efficiency of \spk~and the baselines in terms of training throughput and \benefit, considering scenarios with a single training task as well as multiple tasks on a given-size cluster, without any failures.

\stitle{Comparison with Baselines (single task).} The experiment involves training the GPT-3 model  of different scales on clusters of varying sizes.
Megatron is selected as the baseline because it exhibits significantly higher training throughput compared to other baselines like Oobleck, Varuna, and Bamboo.
To ensure a fair comparison, Megatron is configured with optimal system settings including DP, PP, TP, and others, which are identical to those used for \spk.
Figure~\ref{fig:exp-throughput} presents the training throughput of training the GPT-3 $7B$ model, measured in samples per second.
We observe that \spk~performs on par with Megatron as it introduces no additional overhead to the normal training process and allows all the optimizations implemented in Megatron to be fully utilized.
Figure~\ref{fig:exp-outback-overhead} further validates this claim by comparing the achieved FLOP/s ratio of \spk~with that of Megatron, based on testing the GPT-3 model of varying sizes on a $64$-GPU cluster.

\begin{table}[h]
  \vspace*{1ex}
  \centering
  \caption{The tested cases in multi-task experiments.}
  \label{tab:tasks}
  \fontsize{8}{9}\selectfont
  \setlength{\tabcolsep}{3.5pt}
  \begin{tabular}{|cc|cccccc|}
    \hline
    \multicolumn{2}{|c|}{\textbf{Case}} & \textbf{Task 1} & \textbf{Task 2}  & \textbf{Task 3} & \textbf{Task 4} & \textbf{Task 5} & \textbf{Task 6} \\
    \hline
    \multirow{2}{*}{1} & S.\tnote{a} & 7B & 7B & 7B & 7B & 7B & 7B \\
    & W. & 1.0 & 1.0 & 1.0 & 1.0 & 1.0 & 1.0\\
    \hline
    \multirow{2}{*}{2} & S. & 1.3B & 1.3B & 1.3B & 7B & 7B & 13B \\
    & W. & 1.0 & 1.0 & 1.0 & 1.0 & 1.0 & 1.0 \\
    \hline
    \multirow{2}{*}{3} & S. & 7B & 7B & 7B & 7B & 7B & 7B \\
    & W. & 0.5 & 0.8 & 1.1 & 1.4 & 1.7 & 2.0 \\
    \hline
    \multirow{2}{*}{4} & S. & 1.3B & 1.3B & 1.3B & 7B & 7B & 13B \\
    & W. & 0.5 & 0.8 & 1.1 & 1.4 & 1.7 & 2.0 \\
    \hline
    \multirow{2}{*}{5} & S. & 1.3B & 1.3B & 1.3B & 7B & 7B & 13B \\
    & W. & 2.0 & 1.7 & 1.4 & 1.1 & 0.8 & 0.5 \\
    \hline
  \end{tabular}
  \begin{tablenotes}
    \item[a] \emph{S.} denotes the model size, and \emph{W.} denotes the weight of the task. 
  \end{tablenotes}
  \vspace*{-3ex}
\end{table}

\begin{figure*}[t]
  \centering
  \begin{subfigure}{.7\linewidth}
    \includegraphics[width=\linewidth]{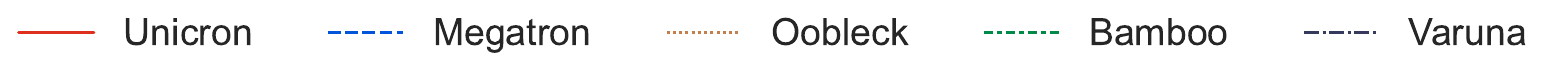}
    \label{fig:exp-e2e-perf-legend}
  \end{subfigure}

  \vspace{-1.5em}
  \begin{subfigure}{.33\linewidth}
    \centering
    \includegraphics[width=\linewidth]{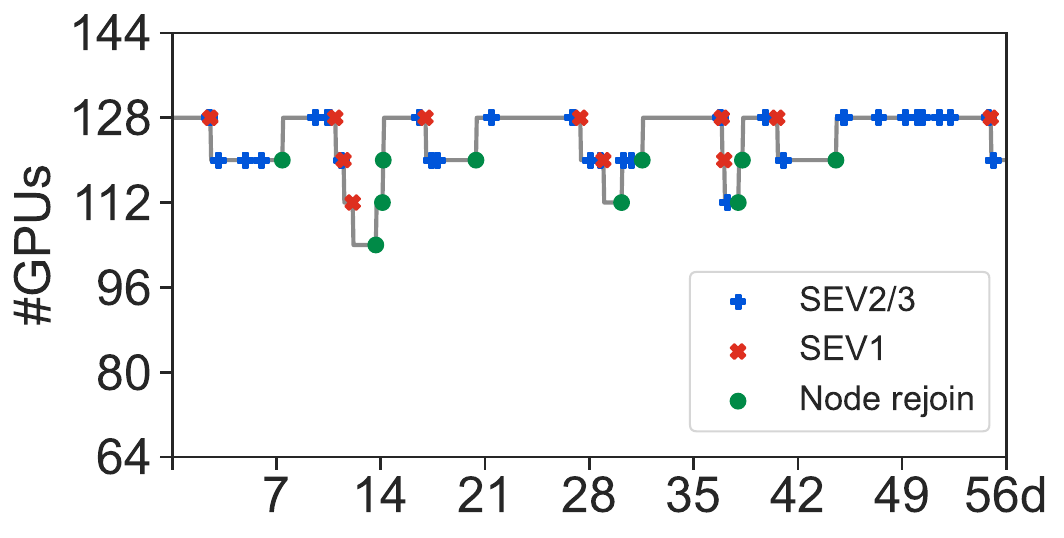}
    \caption{\spk~trace-a.}
    \label{fig:exp-trace-outback-errors-a}
  \end{subfigure}
  \begin{subfigure}{.33\linewidth}
    \centering
    \includegraphics[width=\linewidth]{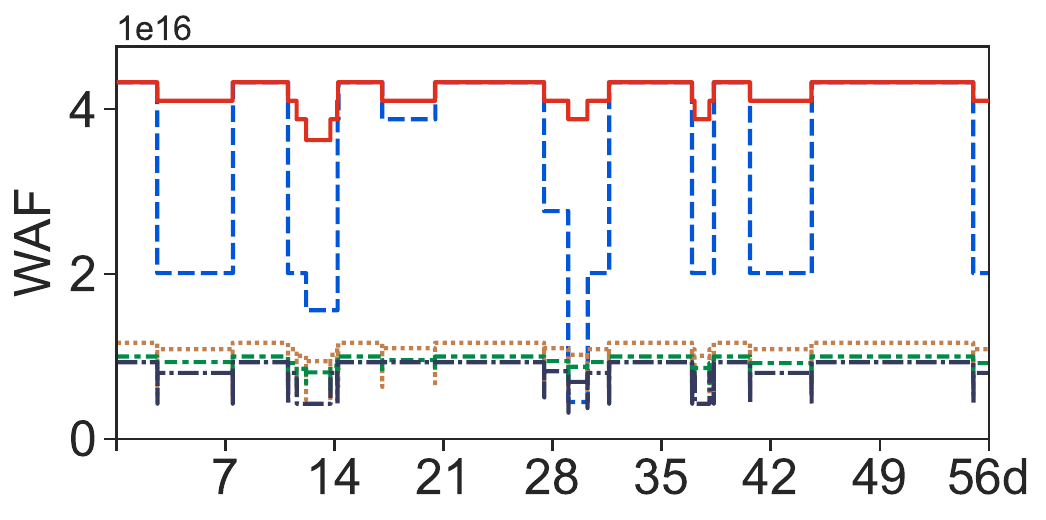}
    \caption{\benefit~(trace-a).}
    \label{fig:exp-trace-outback-errors-a-2}
  \end{subfigure}
  \begin{subfigure}{.33\linewidth}
    \centering
    \includegraphics[width=\linewidth]{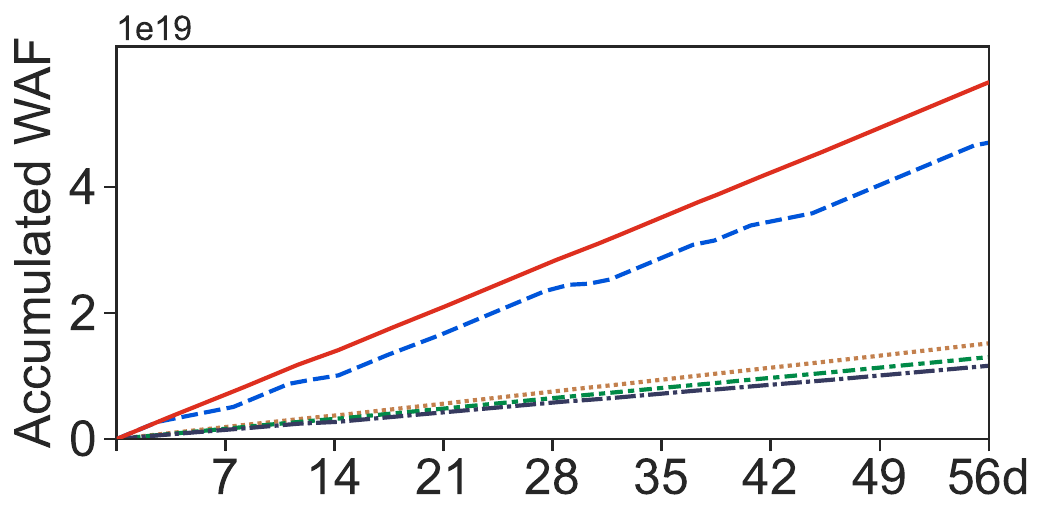}
    \caption{Accumulated \benefit~(trace-a).}
    \label{fig:exp-trace-outback-errors-a-1}
  \end{subfigure}

  \vspace{0.5em}

  \begin{subfigure}{.33\linewidth}
    \centering
    \includegraphics[width=\linewidth]{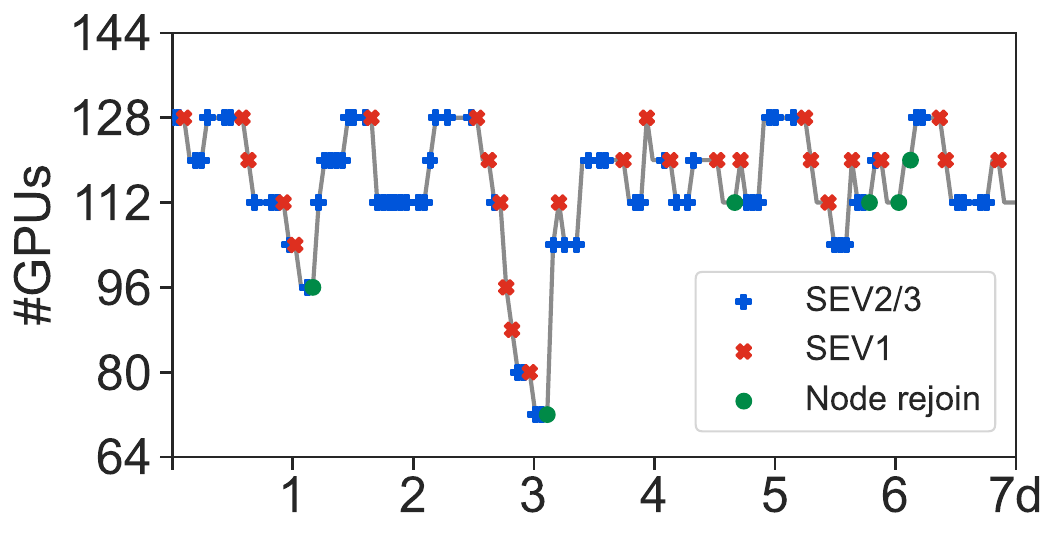}
    \caption{\spk~trace-b.}
    \label{fig:exp-trace-outback-errors-b}
  \end{subfigure}
  \begin{subfigure}{.33\linewidth}
    \centering
    \includegraphics[width=\linewidth]{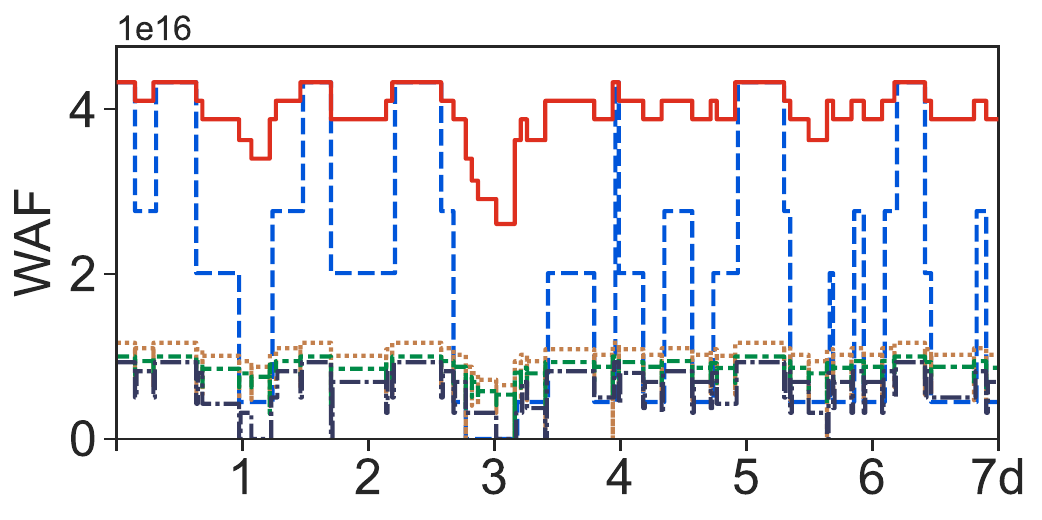}
    \caption{\benefit~(trace-b).}
    \label{fig:exp-trace-outback-errors-b-2}
  \end{subfigure}
  \begin{subfigure}{.33\linewidth}
    \centering
    \includegraphics[width=\linewidth]{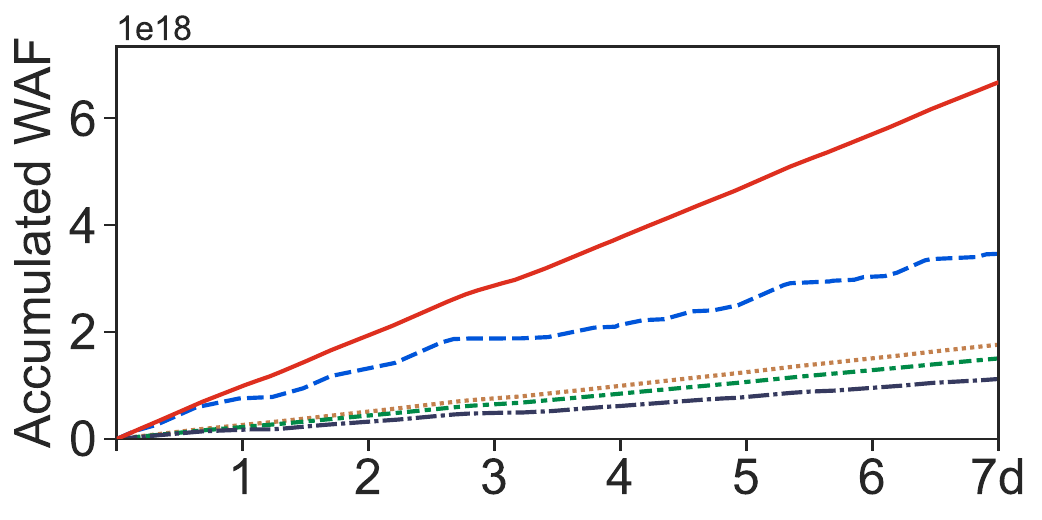}
    \caption{Accumulated \benefit~(trace-b).}
    \label{fig:exp-trace-outback-errors-b-1}
  \end{subfigure}

  \caption{Overall training efficiency (in \benefit~and accumulated \benefit) of \spk~and baselines under different failure traces.}
  \label{fig:exp-e2e-perf}
  \vspace*{-1ex}
\end{figure*}

\stitle{Comparison with Baselines (multiple tasks).} Figure~\ref{fig:exp-benefits} compares \spk~with various baselines when training six tasks on a $128$-GPU cluster, measured in the \benefit~metric of the cluster.
Table~\ref{tab:tasks} summarizes the settings for the five tested cases, which involve training multiple tasks with different sizes and priorities.
Since other systems do not support generating configuration plans for multiple tasks simultaneously, we implemented several baseline strategies for comparison. 
The first baseline, denoted as ``equally'', evenly allocates computing resources among all tasks. The second baseline, denoted as ``weighted'', allocates computing resources based on the weight assigned to each task. 
The third baseline, denoted as ``sized'', allocates computing resources based on the model size.
From the figure, it is evident that \spk~consistently achieves the highest \benefit~across all five cases compared to the baselines. 
This outcome highlights the effectiveness of the configuration plan generated by \spk~in maximizing \benefit, thereby optimizing overall utilization and efficiency.

\subsection{Overall Training Efficiency}
Lastly, we evaluate and compare \spk's overall training efficiency with baselines in various failure trace scenarios.

\stitle{Traces.} We collected a failure trace referred to as \emph{trace-a} in Figure~\ref{fig:exp-trace-outback-errors-a} from real-world logs we collected, as presented in Figure~\ref{fig:errors}.
Note that failure occurrences are considered independently for each GPU or node.
The trace spans a $8$-weeks period, including $10$ \sev{1} failures and $33$ failures of other types. 
On the trace, the x-axis denotes the timeline of the training process, while the y-axis reflects the count of available GPUs in the cluster at any given time.
It is worth mentioning that solely \sev{1} failures lead to a decrease in the count of available GPUs, while other types of failures (\sev{2} and \sev{3}) do not affect this count.
Regarding \sev{1} failures, the time taken for a node to recover and become available once more is determined by a uniform random selection, varying from $1$ to $7$ days.

Additionally, Figure~\ref{fig:exp-trace-outback-errors-b} depicts another trace (referred to as  \emph{trace-b}), which is generated by amplifying the failure frequency of \emph{trace-a} by a factor of $20 \times$. The occurrence of failures during the training process is simulated using a Poisson distribution, which allows for the possibility of multiple failures happening within a given time interval. \emph{trace-b} is designed to simulate a scenario where failures occur more frequently, thus providing a rigorous test of the systems' self-healing capabilities under extreme scenarios.

It spans a duration of $7$ day and records $26$ \sev{1} failures and $80$ other failures.
Accordingly, repaired nodes are re-joining the cluster at a similar rate to maintain a stable resource pool. 
This particular trace serves to replicate an environment in which failures occur more frequently, thus providing a rigorous test of the system's self-healing capabilities.
In our simulation, \sev{1} failures are induced by killing the agent process on the affected node, while \sev{2} and \sev{3} failures are triggered by either raising an exception or by killing/stalling the training process.

\stitle{Workloads and Baselines.} For the evaluation, the workload used is the Case\#5 from Table~\ref{tab:tasks}.
This workload involves the concurrent training of multiple tasks on a cluster with $128$ GPUs.
These tasks vary in size and priority, simulating real-world scenarios where different training jobs are running simultaneously.
Given that the baseline systems we compare against do not possess the capability to generate configuration plans for multiple tasks, we assign  the same initial configuration plan designated for \spk~to these baselines, which is considered optimal according to our proposed model.
However, in the event of a failure, the baseline methods only reconfigure the task directly impacted. Moreover, should a node recover, these methods give precedence to reconfiguring the task that was first affected. 
In contrast, \spk~features a critical advantage over the baseline approaches: it can also reconfigure other tasks if it proves beneficial to do so.

\stitle{Comparison.} In Figure~\ref{fig:exp-e2e-perf}, we present a comparison analysis of the overall training efficiency between \spk~and the baselines.
The efficiency metric employed is the total \benefit~of all tasks within the cluster, measured at successive time points. 
The accumulated \benefit~is also reported to illustrate the overall training efficiency throughout the evaluation period.

The results unequivocally demonstrate that \spk~consistently achieves the highest performance in accumulated \benefit.
Compared with baselines, on \emph{trace-a}, \spk~outperforms Megatron by $1.2\times$, Bamboo by $4.6\times$, Oobleck by $3.7\times$, and Varuna by $4.8\times$ in terms of accumulated \benefit.
It should be noted that Megatron achieves higher performance than Bamboo, Oobleck, and Varuna, which is expected given its integration of various techniques to ensure training efficiency. 
Conversely, Bamboo, Oobleck, and Varuna, with their primary focus on fault tolerance, do not prioritize training efficiency optimizations, resulting in lower performance.

Under \emph{trace-b}, when failure frequency raising, \spk~outperforms Megatron by $1.9\times$, Bamboo by $4.8\times$, Oobleck by $3.8\times$, and Varuna by $5.8\times$ in terms of accumulated \benefit.
All systems experience diminished performance under this trace due to the increased frequency of failures. 
Megatron suffers the most significant reduction as the recovery costs are exacerbated by the frequent failures, and it lacks specific optimizations to mitigate these additional expenses. 
\spk, being built on top of Megatron, is able to fully leverage its optimizations while introducing additional optimizations for efficient self-healing.
As a result, \spk~achieves significant improvements in performance compared to other baselines under this trace. 
These outcome underscores the effectiveness of \spk~in achieving high performance under diverse scenarios, validating its practical applicability.

%% file: sec-conclusion.tex
\section{Conclusion}
\label{sec:conclusion}
This paper introduces the \spk~system as a holistic approach to address the challenges of failure recovery in training large-scale language models.
By incorporating in-band error detection, a dynamic cost-aware plan generation mechanism, and a transition strategy, \spk~minimizes the overall cost of failures across multiple tasks within a cluster. 
As a result, \spk~demonstrates a notable increase in overall training efficiency, with performance gains reaching up to $1.9\times$ that of state-of-the-art solutions.